\documentstyle[preprint,aps,eqsecnum]{revtex}

\begin{document}


\def \mneut#1{\widetilde{M}_{N_{#1}}}
\def \neut#1{{N}_{#1}}
\def \char#1#2{{C}_{#1}^{#2}}

\draft
\preprint{
  \parbox{1.4in}{UM--TH--96--13 \\
  [-0.12in] hep-ph/9609210
}  }

\title{Searching for a Light Stop at the Tevatron}
\author{Gregory Mahlon \cite{GDMemail} and G.L. Kane \cite{GLKemail}}
\address{Department of Physics, University of Michigan \\
500 E. University Ave., Ann Arbor, MI  48109 }
\date{August 30, 1996}
\maketitle
\begin{abstract}
We describe a method to help the search for a light stop squark
($\widetilde{M}_{t} + \widetilde{M}_{LSP} < {m}_{t}$) at the Fermilab 
Tevatron. Traditional search methods rely upon a series of stringent 
background-reducing cuts which, unfortunately, leave very few signal 
events given the present data set.  To avoid this difficulty, we 
instead suggest using a milder set of cuts, combined with a 
``superweight,'' whose purpose is to discriminate between signal and 
background events.  The superweight consists of a sum of terms, each 
of which are either zero or one.  The terms are assigned event-by-event 
depending upon the values of various observables.  We suggest a method 
for choosing the observables as well as the criteria used to assign 
the values such that the superweight is ``large'' for the 
supersymmetric signal and ``small'' for the standard model background.  
For illustration, we mainly consider the detection of stops coming 
from top decay, making our analysis especially relevant to the 
$W$ + 2 jets top sample.  
\end{abstract}
\pacs{14.80.Ly, 12.60.Jv, 14.65.Ha}


\section{Introduction}

There has been recent activity in the area of
weak-scale supersymmetry, spurred on by a number of suggestive
experimental results. 
First, there is the single 
$ee\gamma\gamma + 
\thinspace\thinspace{\not{\negthinspace\negthinspace E}}_T$ event
observed by CDF\cite{eegamgam}.  This particular event does
not seem to have a Standard Model interpretation.
Also, in supersymmetry, the $Z\rightarrow b\bar{b}$ rate 
($R_b =  \Gamma(Z \rightarrow b \bar b) /
\Gamma(Z \rightarrow {\rm hadrons})$),
the value of $\alpha_s$ extracted from $\Gamma_Z$, and
the branching ratio for $b\rightarrow s\gamma$ are all
affected by loop diagrams containing charginos and stop squarks.
At present \cite{Warsaw}, all three of these quantities are 
1.5--2.0$\sigma$ from their Standard Model predictions,
each in precisely the directions expected from supersymmetry
\cite{Warsaw,Shifman,RbA,KaneKoldaWells,CDM,Global} 
if there is a light stop squark.

Remarkably, these  
experimentally independent ``mysteries'' can all be
explained by a single reasonably well-determined set of parameters
within the framework of weak scale supersymmetry.  
The $ee\gamma\gamma + 
\thinspace\thinspace{\not{\negthinspace\negthinspace E}}_T$ event 
has a natural interpretation
in terms of selectron pair production.
Two different scenarios  are possible, depending upon whether
the lightest supersymmetric particle (LSP) is a gravitino
\cite{selectronA,selectronB}
or Higgsino-like neutralino\cite{selectronA}.  
The neutralino and chargino parameters suggested by the
second scenario overlap with the values required to 
account for the $R_b$ difference\cite{RbA},
provided that one of the stop squark eigenstates is
light ($\widetilde{M}_{t} \alt m_W$).

One might be concerned that such a  low stop mass
would have undesirable side-effects.  Indeed, an immediate 
consequence\cite{KaneKoldaWells} is that the decays
\begin{equation}
t \rightarrow \tilde{t}\neut{i}
\label{TopToStopDecay}
\end{equation}
where $\tilde{t}$ is the lighter of the two stop mass eigenstates
and $N_i$ is a kinematically accessible 
neutralino mass eigenstate should 
occur with a total branching ratio in the 
neighborhood of 50\%.  
The consequences of this depend on how the stop
decays.  
When at least one chargino is light enough, the decay
\begin{equation}
\tilde{t} \rightarrow C_i b
\label{OtherDecay}
\end{equation}
dominates~\cite{StopPhenom}.  
In this case, the subsequent chargino decay
to a fermion-antifermion pair plus neutralino leads to
a $\neut{1}\neut{1}f \bar{f}' b$ final state whenever
a top undergoes the decay~(\ref{TopToStopDecay}).  Since the
final state for the SM decay is identical except for the (invisible)
neutralinos, there is potential for both direct stop decays as
well as top to stop decays to mimic ordinary top decays.
This possibility has been investigated
by several authors\cite{StopPhenom,BST,Lopez,Abraham,Sender}.
In particular, we note that 
Sender~\cite{Sender} finds that
models with a ``large'' ${\cal B}(t \rightarrow \tilde{t}\neut{1})$
have not been ruled out by Tevatron data, provided the stop
decays according to~(\ref{OtherDecay}).
This scenario, although interesting, is not the focus of this
paper.  Instead, we wish to examine the situation where
the one-loop decay
\begin{equation}
\tilde{t} \rightarrow \neut{1} c
\label{OneLoop}
\end{equation}
dominates, which happens when the decay~(\ref{OtherDecay}) is 
kinematically forbidden~\cite{FCNC}.  In this case, 
a top undergoing the decay~(\ref{TopToStopDecay}) would 
produce a $c\neut{1}\neut{1}$ final state and
be effectively invisible to standard searches.
Two independent analyses appropriate to this case 
have been performed~\cite{Sender,Yuan}
which conclude that ${\cal B}(t \rightarrow X)$
where $X\ne Wb$ is at most 20--25\%.
However, neither analysis accounts for the possibility
that supersymmetry can lead to additional sources of 
top quarks, without the need for stop decays masquerading as
top decays~\cite{XtraTops}.  For example, if the gluino
is lighter than the other (non-stop) squark flavors, but
heavier than ${m}_{t} + \widetilde{M}_{t}$, then it decays 
exclusively via~\cite{StopPhenom}
\begin{equation}
\tilde{g} \rightarrow t \tilde{t}^{-}, \bar{t}\tilde{t}^{+},
\label{GluinosToTheRescue}
\end{equation}
making the production of gluinos a source 
of top quarks~\cite{XtraTops,Kon}.
In fact, the authors of Ref.~\cite{XtraTops} argue that there is 
indirect evidence for the decays~(\ref{TopToStopDecay}),
(\ref{OneLoop}), and (\ref{GluinosToTheRescue}) in the 
Fermilab data on top rates and distributions.

In light of the indirect hints at weak scale supersymmetry, it
is important to take every opportunity to
obtain some direct evidence that nature is indeed supersymmetric, 
or, to show that it is not.
A discussion of search strategies for the direct production
of stop pairs at the Tevatron already exists
in the literature\cite{BST}.  
The authors of Ref.~\cite{BST} claim that a stop
squark with a mass of up to about 100 GeV
should be visible at the Tevatron in the $\ge2$ jets plus 
missing transverse energy channel given 100 pb$^{-1}$ of data.
Nevertheless, 
we feel that it is beneficial to augment the direct search
with a search for stops coming from top decay:
observation of a signal in both channels would greatly 
boost the case for SUSY.
To this end, we present a method that may facilitate the
search for top to stop decay at the Fermilab Tevatron.
Our method consists of defining a 
superweight $\widetilde{\cal X}$ whose
function is to discriminate between signal and background events.
The superweight is constructed from various observables 
in the events so that it is ``large'' for the signal events 
and ``small'' for the background.
We will illustrate the superweight method using the case
of the stops coming from top quark decays;  it can also
be applied to stop pair production.  Although the required 
analysis is not easy, it should be possible to determine
directly whether about half of all tops indeed decay to stops.
Our goal in this paper is to help this process.

The authors of Ref.~\cite{Yuan} have also looked at the
problem of searching for stops from top decay at the Tevatron.
However, they employ the traditional method of cutting
only on the kinematic observables in the event.
As a result, their signal efficiencies are rather low 
(6\%--8\%).  Also, they do not include the possibility of
SUSY-induced $t \bar t$ production.  Consequently, they conclude
the prospects for observing a signal, if present, are
most promising at an {\it upgraded}\ Tevatron.  
As we shall see, the superweight method allows us to reduce
the backgrounds to a level similar to that of Ref.~\cite{Yuan},
but with efficiencies as high as 16\%, providing for
the possibility of finding a signal in the {\it current}\ data set.

\begin{figure}[h]
\vspace*{15cm}
\includegraphics{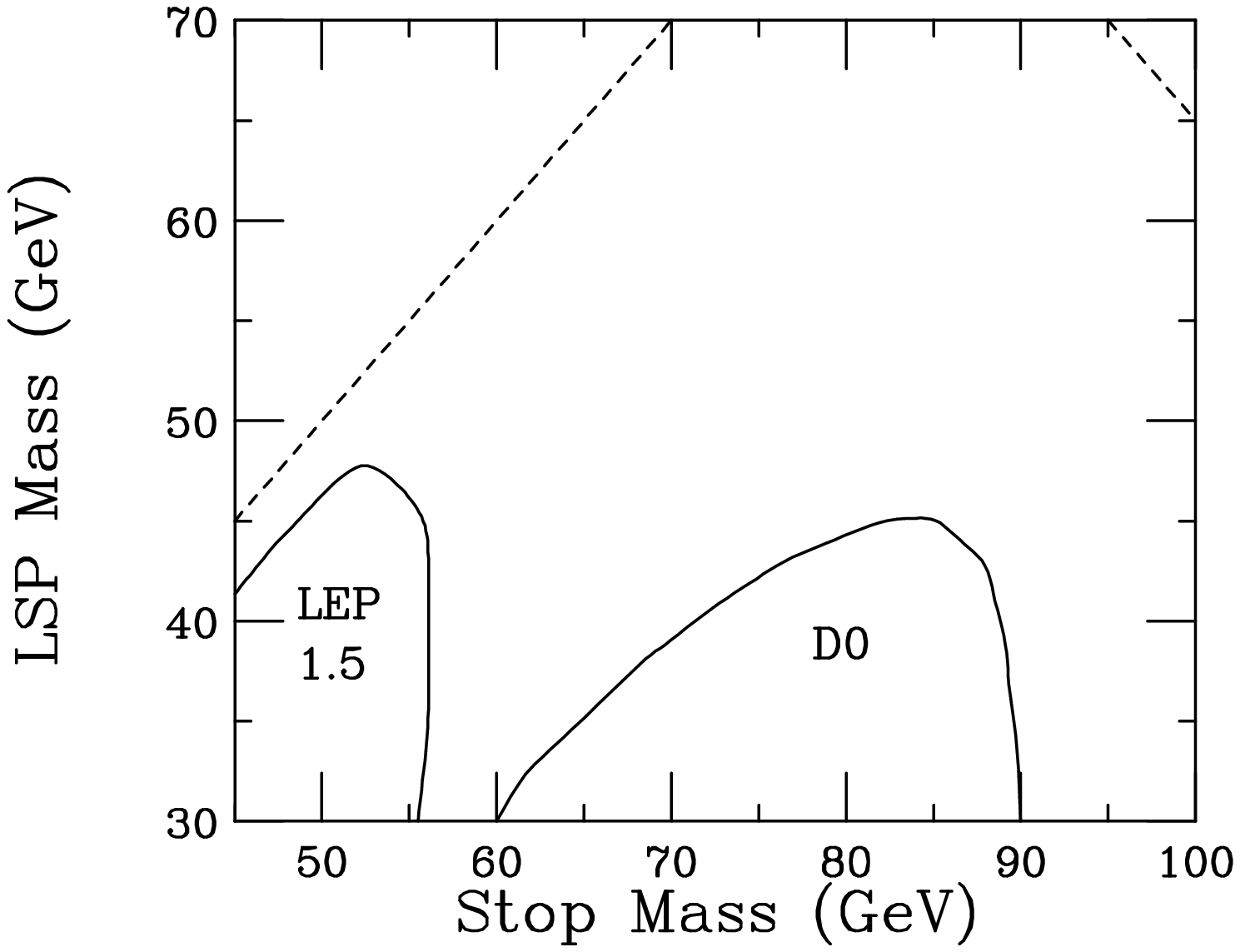}
\vspace{2.0cm}

\caption[]{Regions of the $\widetilde{M}_{t}$-$\mneut{1}$ mass
plane excluded by D0 \cite{D0stop}
and LEP 1.5 \cite{LEPstop}.  The area above the dashed line
is kinematically forbidden.
}
\label{exclusion}
\end{figure}

The D0 experiment at Fermilab as well as the various LEP 
experiments have reported limits based upon searches
for the pair production of stop squarks~\cite{D0stop,LEPstop}
(see Fig.~\ref{exclusion}).  
In the event that the current
run at LEP finds a stop signal, the confirmation process
could be greatly aided by the Tevatron data, depending
upon the stop and LSP masses.  On the other hand, even if
LEP sees nothing, there is still a significant region in the
stop-LSP mass plane to which the Tevatron is sensitive and
which will not have been excluded by LEP.

The remainder of the paper is organized as follows:
in Sec.~\ref{Features} we
briefly examine  the generic features  of SUSY models
hinted at by the data,
and determine the experimental signature we will concentrate on.
Sec.~\ref{SUwgt} contains a general discussion of the superweight
and the methods by which it is constructed.
We discuss the detection of the decay $t \rightarrow \tilde{t}\neut1$
in Sec.~\ref{BasicSignal}, within the framework of a simplified
model where no other neutralinos are light enough to be
produced, and where only SM $t \bar t$ production mechanisms
are considered.  Such an analysis is appropriate for any
SUSY model which contains the decays~(\ref{TopToStopDecay})
and~(\ref{OneLoop}), whether or not there are extra sources
of top quarks.
We expand our discussion to include the other neutralinos
and SUSY $t \bar t$ production mechanisms in Sec.~\ref{FullSignal}.
Finally, Sec.~\ref{CONCLUSIONS}
contains our conclusions;  Fig.~\ref{Nevents} summarizes our
results.


\section{Stops from Top Decay} \label{Features}

\subsection{A SUSY Model}

In this section we flesh-out the 
supersymmetric scenario described in the introduction.
Specifically, the picture implied by 
Refs.~\cite{RbA,selectronA,XtraTops} contains a Higgsino-like
neutralino $\neut1$ with a mass in the 30 to 55 GeV range,
a light stop squark with a mass in the 45 to 90 GeV range,
a gluino with a mass in the 210 to 250 GeV range, 
and $\tilde{u},\tilde{d},\tilde{s},\tilde{c}$ squarks
in the 225 to 275 GeV range.  The ``heavy'' stop eigenstate
as well as both $\tilde{b}$ squark eigenstates 
may be heavier.
We further assume that the ``light'' stop eigenstate
is lighter than the charginos, and that the gluino is lighter 
than the squarks (except for $\tilde{t}$).  
The stop and the lighter chargino could be 
approximately degenerate; we ignore such a complication here.
We take the top quark mass 
to be 163 GeV.

With these masses and couplings, the decays and branching ratios
relevant to our study are
\begin{equation}
{\cal B}(\tilde{q} \rightarrow q\tilde{g}) \sim 25\% \hbox{--} 75\%
\label{SquarkDecay}
\end{equation}
\begin{equation}
{\cal B}(\tilde{g} \rightarrow t\tilde{t}^{-}) =
{\cal B}(\tilde{g}\rightarrow\bar{t}\thinspace\tilde{t}^{+}) = 50\%
\end{equation}
\begin{equation}
{\cal B}(t \rightarrow Wb) \sim 
{\cal B}(t \rightarrow \tilde{t}\neut{i}) \sim 50\%
\label{ThreeThree}
\end{equation}
\begin{equation}
{\cal B}(\tilde{t} \rightarrow c\neut{1}) \sim 100 \%.
\end{equation}
The large variation in the branching ratio for squark decays
is a consequence of the relatively small phase space available for
producing gluinos; hence, 2-body decays to the electroweak 
superpartners are able to compete effectively with the strong decays.
The gluinos, however, have no other 2-body decay modes:  if
top-stop is open, it dominates.

As indicated by~(\ref{ThreeThree}), the total branching
ratio for top to all of the kinematically accessible neutralinos
is about 50\%.  
In these models, $N_1$ is the LSP and
assumed stable.  The interpretation of the $ee\gamma\gamma$
which inspired our closer examination of this particular
region of parameter space requires
\begin{equation}
{\cal B}(\neut{2} \rightarrow \neut{1}\gamma) > 50\%.
\end{equation}
In principle, one could look for this photon as an aid
in selecting events with $t \rightarrow\tilde{t}\neut{2}$.
However,
because of the large photino content of the $N_2$, 
its production in top decays is suppressed compared to
$N_1$ or $N_3$.  So rather than concentrate on 
a small fraction of events, we make no attempt to 
identify the photon in our study, and instead allow it
to mimic a jet.
Quite often, $N_3$ is also light enough to be produced 
by decaying tops.
Its decays are
more complicated:  if the sneutrinos happen to be light
enough to provide a 2-body channel, then  $\tilde{\nu}\nu$ is favored;
otherwise, the 3-body
decays $\neut{1}f\bar{f}$ where $f$ is a light fermion
dominate.   The net result is of all of this is that
allowing the top to decay to SUSY states other than $N_1$
simply adds additional (relatively) soft jets to the final
state.

\subsection{The Supersymmetric Signal}

Within the supersymmetric scenario proposed in Ref.~\cite{XtraTops},
there are several different production mechanisms for top
quarks, and hence many different final states which must be 
considered.

For the top pairs produced by the usual SM processes,
we end up with mainly three different final states, depending upon
the way in which they decay.
Firstly, both tops could decay to $Wb$, according to the Standard Model.
In this case, the final state consists of
2 leptons, 2 jets, and missing $p_T$ (dilepton);
1 lepton, 4 jets, and missing $p_T$ ($W$ + 4 jets); or
6 jets (all jets).
Secondly, one top could decay to $Wb$, and the other 
to $\tilde{t}\neut1$.
In this case, the final state consists of 
1 lepton, 2 jets, and missing $p_T$ ($W$ + 2 jets);
or 4 jets and missing $p_T$ (missing + 4 jets).
Finally, both tops could decay to $\tilde{t}\neut1$.  In this
case the final state consists entirely of 2 jets and missing 
$p_T$ (missing + 2 jets).  This last final state is identical
to that of direct stop pair production, which is considerably
more difficult because of the large 
Standard Model multijets background.
Of the remaining final states coming
from supersymmetric sources, the $W$ + 2 jets mode is 
the most promising, and the one we will discuss in detail.
In our illustrations, we will describe the situation where
the top decays to a supersymmetric final state, and 
the antitop decays according to the Standard Model:
\begin{equation}
p \bar p \rightarrow t \bar t \rightarrow c \neut1 \neut1 \bar b
\ell^{+} \bar\nu_\ell.
\label{TopToStopSignal}
\end{equation}
The presence of the charge-conjugated process is always implicitly 
assumed, and is included in all of the rates reported below.

In addition to~(\ref{TopToStopSignal}), 
we must consider the effect of 
top quarks arising from gluino and squark decays, which,
as argued in \cite{XtraTops}, must be present in significant
numbers if the non-$Wb$ top quark branching ratio is to be
as large as is typical for a light stop.  Top quarks produced
in this manner are accompanied by extra jets.  Consider
first the pair production of gluinos.  Both gluinos will decay
to a top and a stop.  The tops then decay
as described above, and the stops each yield a charm jet
and a neutralino.
Thus,
gluino pair production leads to the same final states as top
pair production, but with two additional charm jets and
additional missing energy.
Likewise, for the chain beginning with a squark,
we pick up an additional jet from the decay~(\ref{SquarkDecay}).
For a summary of conventional gluino physics at FNAL, see
Ref.~\cite{Haber}.

\begin{figure}[h]
\vspace*{15cm}
\includegraphics{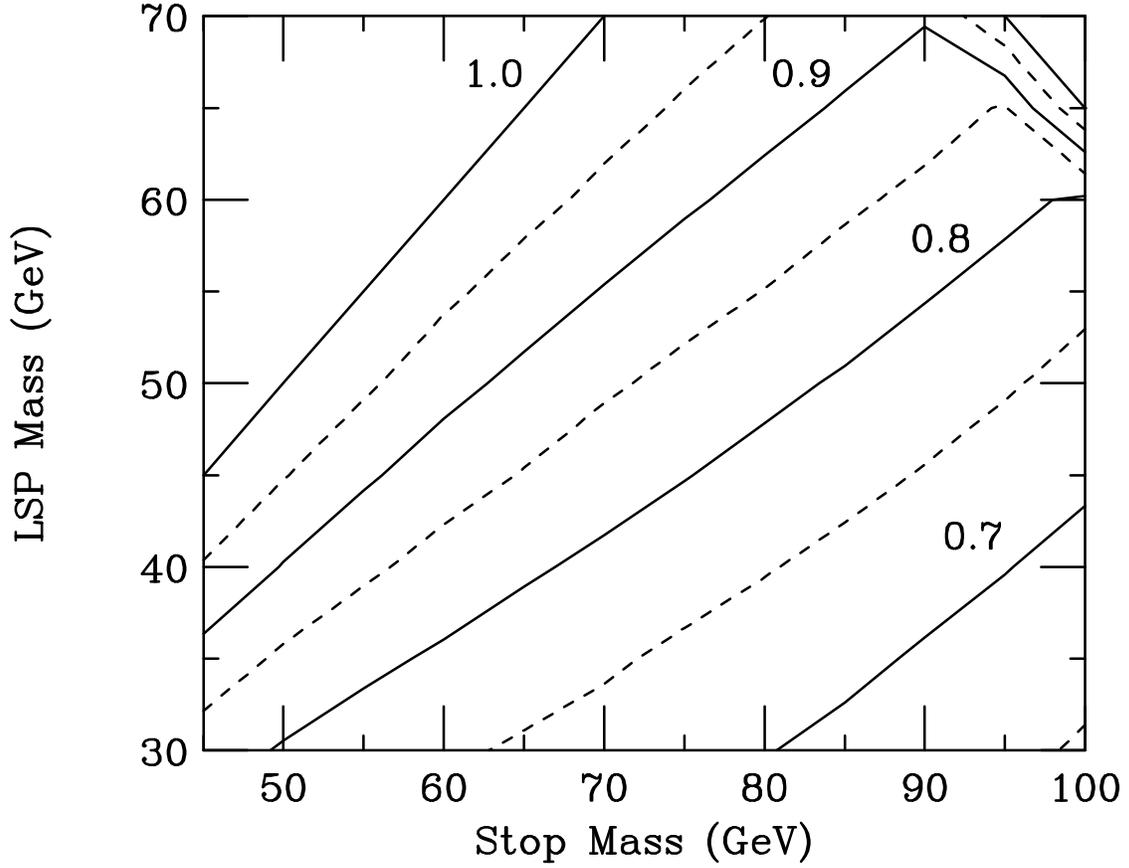}
\vspace{2.0cm}

\caption[]{Fraction of 
$t \bar t \rightarrow c \neut1 \neut1 \bar b \ell^{+} \bar\nu_\ell$
events surviving all of the
cuts in Table~\ref{TopToStopCuts}
for which the $\bar{b}$ quark becomes
the highest $E_T$ jet ($j_1$), as a function of the stop and LSP
masses.  The contour labeled 1.0 is also the kinematic limit.
}
\label{BfractLEGO}
\end{figure}

It is useful to examine some kinematical consequences of the
scenario we have proposed.
Consider first purely SM $t \bar t$ production at the Tevatron,
which takes place
relatively close to threshold.
We would expect
the ordering in $E_T$ of the $\bar b$ and $c$ jets 
to reflect fairly accurately 
the relative sizes of the $t$-$W$
and $\tilde{t}$-$\neut1$ mass splittings.   For the range of masses
we consider here, 
$ m_t - m_W > \widetilde{M}_{t} - \mneut1.  $
Hence, the highest $E_T$ jet should come
from the $\bar b$ quark most of the time.  
Our simulations confirm this, with the $\bar b$ quark
becoming the leading jet more than 70\% of the time over most
of the range of SUSY masses examined.
Fig.~\ref{BfractLEGO} shows the results for the kinematically
allowed masses in the ranges 
$30 {\rm \enspace GeV} < \mneut{1} < 70 {\rm \enspace GeV}$,
$45 {\rm \enspace GeV} < \widetilde{M}_{t} < 100 {\rm \enspace GeV}$. 
The situation is only slightly worse when we add squark and gluino
production.  The additional jets from the cascade decays down
to top are rather soft, given the relatively small mass splittings
involved.  Thus,  in the all of the cases we examine here, 
the identification of the $\bar b$ parton with the leading
jet is a reasonably good one.

Because the jets coming from gluino and squark decays are
relatively soft,
we will organize our results around the premise that the
process~(\ref{TopToStopSignal})
is the framework about which
the complications from such decays are relatively small perturbations.
That is, we will first describe the situation as if the
only processes going on are the SM backgrounds plus decays
of top to stop, parameterized by the stop mass, the LSP mass,
and the branching ratio ${\cal B}(t\rightarrow\tilde{t}\neut1)$ 
(Sec.~\ref{BasicSignal}).
Then, we will expand our consideration to include a full-blown
SUSY model where
additional top quarks are being produced by decaying gluinos
(Sec.~\ref{FullSignal}).  As we shall see, 
the same superweight derived under the simplifying assumptions
works well in the more realistic environment.


\section{The Superweight} \label{SUwgt}

We now describe a procedure which may be employed to 
construct a quantity we call the ``superweight'' out
of the various observables associated with a given process.
In principle, this procedure may be used to differentiate
between signal and background in a wide range of processes,
although we will concentrate on the detection of the 
decay~(\ref{TopToStopDecay}).

For each event in the data sample passing our selection criteria
(correct number and stiffness of jets, sufficient missing
energy, correct number of leptons, {\it etc.})
we define a superweight $\widetilde{\cal X}$ by a sum of the form
\begin{equation}
\widetilde{\cal X} = \sum_{i=1}^{N} {\cal C}_i
\label{SUwgtGeneric}
\end{equation}
where the ${\cal C}_i$'s evaluate to 0 or 1 depending upon whether
or not some given criterion is satisfied.  
The number of terms $N$ in the sum defining $\widetilde{\cal X}$ is
arbitrary:  one should use as many terms as there 
are ``good'' criteria.
One could
consider a more general form including separate weighting factors
for the components and continuous values for the ${\cal C}_i$'s
(as in a full-blown neural net analysis).  However, our intent
is to search for new particles over some range of masses
and couplings.  In such  a situation, too much refinement
could narrow the range of parameters to which the superweight
is a good discriminant between signal and background.
Furthermore,
the components appearing in (\ref{SUwgtGeneric}) 
are easily given a physical interpretation, which guides
us in the optimization of the ${\cal C}$'s.

Let us consider a criterion of the form
\begin{equation}
{\cal C} =  \cases{1, & if $ {\cal Q} > {\cal Q}_0$; \cr
                0, & otherwise, \cr}
\label{Crit}
\end{equation}
where ${\cal Q}$ is some measurable quantity associated with
the event, and ${\cal Q}_0$ is the cut point\cite{OtherWay}.
Although we will refer to ${\cal Q}_0$ as a cut point, we 
don't actually cut  events from the sample which
have ${\cal C}=0$.
Note that~(\ref{Crit}) implies that the value of ${\cal C}$
averaged over the entire sample is exactly the fraction of 
the cross section satisfying the constraint ${\cal Q} > {\cal Q}_0$.
A ``good'' superweight component should have the property that
its average for Standard Model events is much less than its average
for SUSY events.  That is, we want
\begin{equation}
\Delta{\cal C} \equiv \langle{\cal C}\rangle_{\rm SUSY}
                  -\langle{\cal C}\rangle_{\rm SM}
\label{IFdiff}
\end{equation}
to be as large as possible.

So, to develop a new superweight, one should first devise a set
of cuts to produce a data set 
where the number of background events versus the number of
signal events is reasonable (S:B of order 1:4, say).
Next, separate Monte Carlos of
both the signal and main backgrounds should be run, in order
to generate plots of $\Delta{\cal C}$ as a function of ${\cal Q}_0$.
The physical interpretation of $\langle{\cal C}\rangle$ as the
fraction of events satisfying ${\cal Q} > {\cal Q}_0$ may be used as a
guide when deciding which ${\cal Q}_0$'s are worth investigating.
For each value of the new physics parameters, there will be
an ideal value of ${\cal Q}_0$ for which $\vert\Delta{\cal C}\vert$
is maximal.  A good superweight component should not only
have a ``large'' value of $\vert\Delta{\cal C}\vert$, but the
corresponding value of ${\cal Q}_0$ at that point should be reasonably 
stable over the entire parameter space to be investigated.

An issue that arises concerns the question of correlations
among the ${\cal C}_i$'s.  Our philosophy in this respect is
to evaluate the effectiveness of the superweight in terms of
how well it separates the signal from the background, {\it i.e.}\ what
is the purity of an event sample with a certain minimum
superweight?  Thus, while we avoid using two ${\cal C}_i$'s whose
values are 100\% correlated (on the grounds that doing
so is no more beneficial than using only one of the two),
we don't worry about using partially correlated ${\cal C}_i$'s.
The main effect of correlations among the ${\cal C}_i$'s is
that the overall performance of the sum of the ${\cal C}_i$'s will
be less than what is  implied by 
considering  the ${\cal C}_i$'s individually.
Thus, to evaluate the effectiveness of a given superweight definition,
one should compare the predicted distributions 
in $\widetilde{\cal X}$ for the signal and background.

We now give an example of the steps used to determine
one of the superweight elements for the
$t \rightarrow\tilde{t}\neut1$ search method described in
detail in Sec.~\ref{BasicSignal}.  To begin, we take a moment to 
recall the definition of the transverse mass.
Given particles
of momenta $P$ and $Q$, the transverse mass of the pair
is defined by
\begin{equation}
m_T^2(P,Q) = 2 P_T Q_T [ 1 - \cos {\it\Phi}_{PQ}],
\label{transX}
\end{equation}
where $P_T \equiv \sqrt{P_x^2+P_y^2}$ and ${\it\Phi}_{PQ}$
is the azimuthal opening angle between $P$ and $Q$.
An important feature of the transverse mass is that if the
particles $P$ and $Q$ were produced in the decay of some
parent particle $X$, then the maximum value of $m_T(P,Q)$ is
precisely the mass of $X$.

\begin{figure}[h]
\vspace*{15cm}
\includegraphics{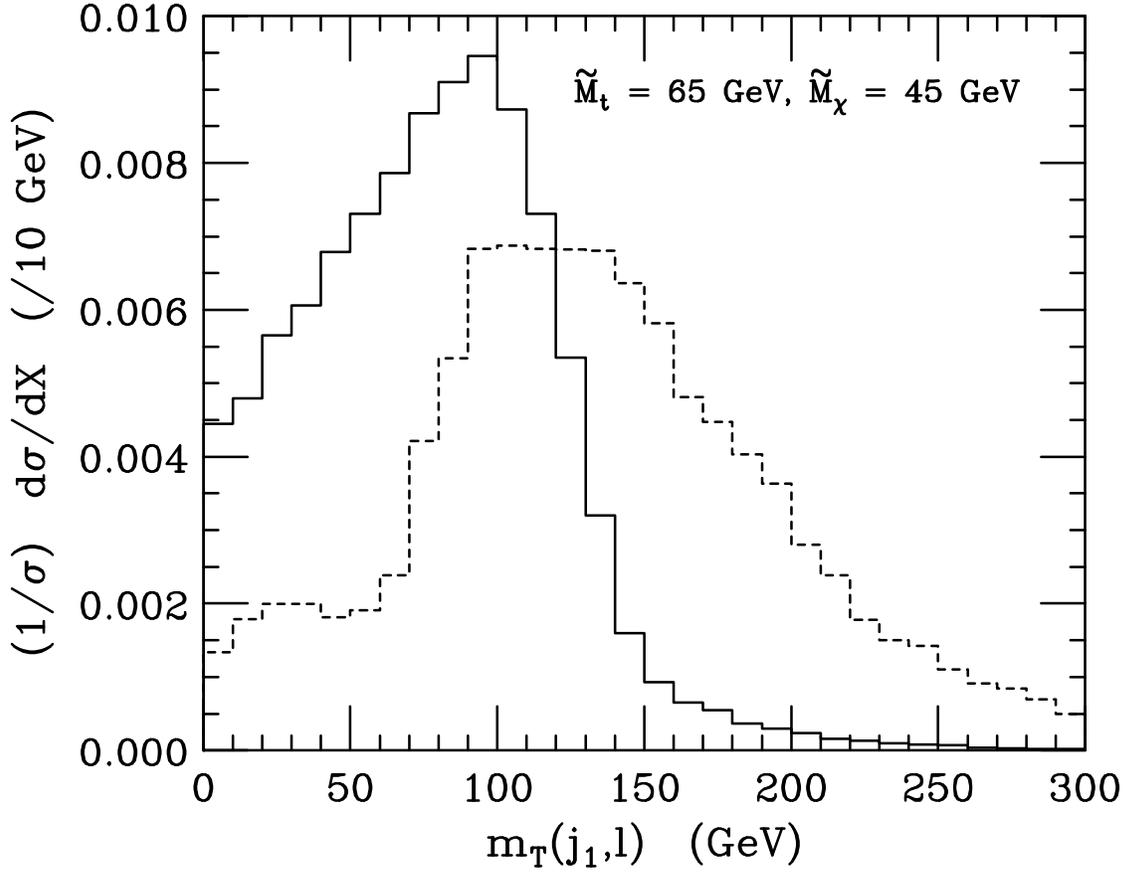}
\vspace{1.0cm}

\caption[]{Differential distribution of the $j_1\ell$ transverse
pair mass for Standard Model $W$ + 2 jets production (dashed) and 
$t \bar t \rightarrow c \neut1 \neut1 \bar b \ell^{+} \bar\nu_\ell$ 
(solid).  We take 
$\widetilde{M}_{t} = 65 {\rm \enspace GeV}$ and 
$\mneut1 = 45 {\rm \enspace GeV}$.
Only events passing all of the cuts in Table~\ref{TopToStopCuts}
are included.
Both histograms have been normalized to unit area.}
\label{SuperPlot}
\end{figure}

\begin{figure}[h]
\vspace*{15cm}
\includegraphics{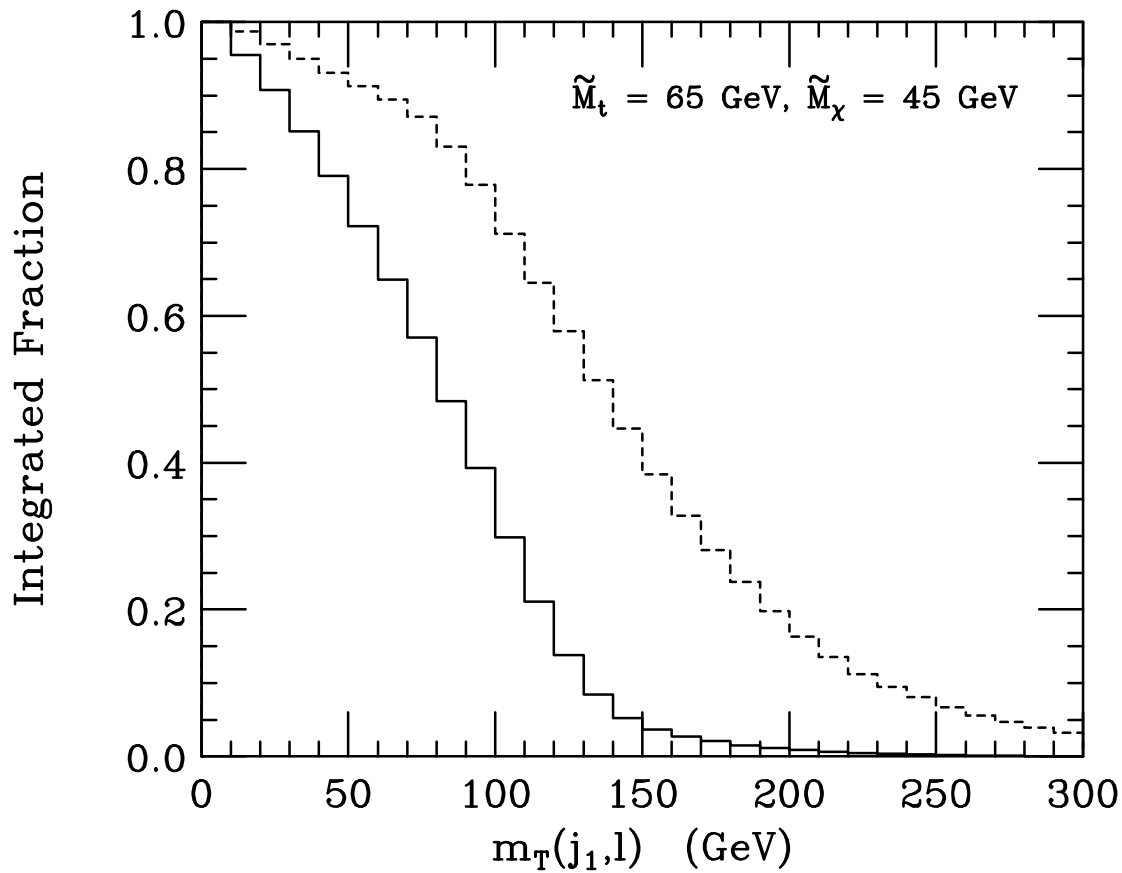}
\vspace{1.0cm}

\caption[]{Fraction of events with a $j_1\ell$ 
transverse pair mass above
$m_T(j_1,\ell)$ for Standard Model $W$ + 2 jets production (dashed) and
$t \bar t \rightarrow c \neut1 \neut1 \bar b \ell^{+} \bar\nu_\ell$
(solid).
We take 
$\widetilde{M}_{t} = 65 {\rm \enspace GeV}$ and 
$\mneut1 = 45 {\rm \enspace GeV}$.
Only events passing all of the cuts in Table~\ref{TopToStopCuts}
are included.}
\label{AAPlot}
\end{figure}

\begin{figure}[h]
\vspace*{15cm}
\includegraphics{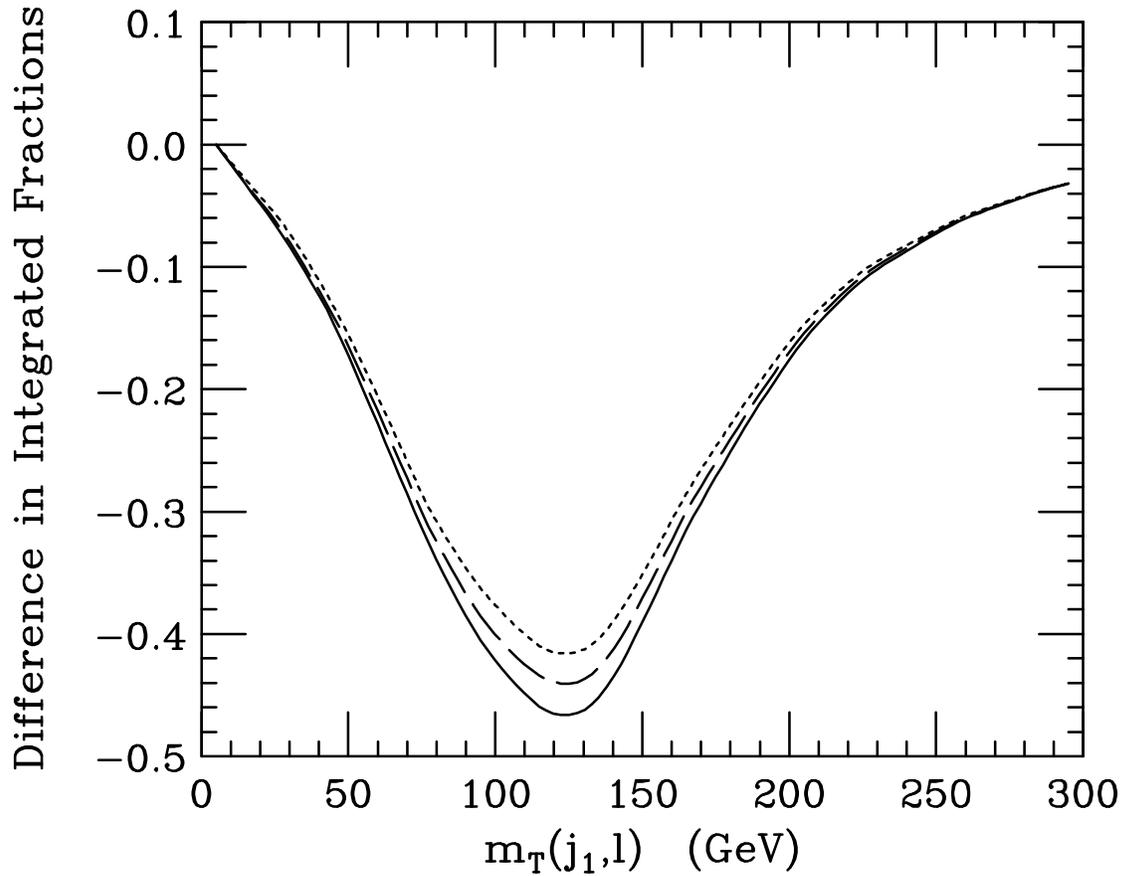}
\vspace{1.0cm}
\caption[]{Difference in integrated fractions, as defined by
equation (\ref{IFdiff}), for the observable $m_T(j_1,\ell)$.
The mass values $(\widetilde{M}_{t},\mneut1)$ in GeV for each line are:
$(50,40)$ solid,
$(65,45)$ dashed,
$(85,50)$ dotted.
}
\label{DiffPlot}
\end{figure}

As already discussed in Sec.~\ref{Features},
the signal~(\ref{TopToStopSignal})
for top to stop 
appears in the detector as a charged lepton, 
2 jets, and missing energy.
The largest background turns out to be
the Standard Model production of a $W$ plus 2 jets, so we determine our
superweight criteria using that background.
Furthermore, we know that for signal events, the
leading jet is usually from the $\bar{b}$ quark in the 
$\bar{t}\rightarrow W^{-} \bar{b}$ decay.
Consequently, most of the time the leading jet and
the charged lepton 
should reconstruct to no more than the top quark mass
(some energy and momentum is carried away by the unseen neutrino).
This suggests
an upper limit on the value of $m_T(j_1,\ell)$, which may
violated at least some of the time by ordinary $W$ plus 2 jet 
events.
In Fig.~\ref{SuperPlot}, we show the differential cross section
in $m_T(j_1,\ell)$
for both the signal and the background, as determined 
from VECBOS\cite{Vecbos} (relevant details of our simulations
will be discussed in Sec.~\ref{BasicSignal}).
Note that for the signal there is the expected sharp drop-off
for large values of $m_T(j_1,\ell)$.
In Fig.~\ref{AAPlot} we show the
fraction of events with a $j_1\ell$ transverse 
mass above $m_T(j_1,\ell)$,
as a function of $m_T(j_1,\ell)$.  
The individual magnitudes of these two
curves are not critical in making a good superweight component,
but rather the difference in these two curves, which is plotted
in Fig.~\ref{DiffPlot} not only for the masses used in 
Figs.~\ref{SuperPlot} and~\ref{AAPlot}, but also for two additional
values as well.  The presence of a dip ranging in depth
from about $-0.4$ to $-0.5$ in the vicinity
of $m_T(j_1,\ell) = 125 {\rm \enspace GeV}$ for each of the 
masses used suggests that 
this is indeed a worthwhile superweight element, and that
the criterion should read
\begin{equation}
{\cal C} = \cases{1, & if $ m_T(j_1,\ell) < 125 {\rm \enspace GeV}$; \cr
                  0, & otherwise. \cr}
\end{equation}
The key quantities to look
for in this evaluation were approximate stability in peak (dip)
position and ``large'' magnitude for the peak for the range of
parameters to be investigated.  Narrowness of the peak is not
a requirement.  In fact, a broad peak is better, since then the
exact placement of the cut point is unimportant.  
Note also that since we are exploiting the difference in the
{\it shapes}\ of the signal and background distributions, 
there is no reason we can't use an observable both for cutting
and in the superweight.  For example, even after requiring
a minimum missing transverse momentum, we can (and do) still use a 
superweight criterion based on the
shapes of the  missing transverse momentum
distributions for the surviving events.


\section{The Process $\noexpand\lowercase{p} 
\bar{\noexpand\lowercase{p}} \rightarrow 
\noexpand\lowercase{t} \bar{\noexpand\lowercase{t}} 
\rightarrow \noexpand\lowercase{c} 
\neut1 \neut1 \bar{\noexpand\lowercase{b}} 
\ell^{+} \bar\nu_\ell$} \label{BasicSignal}

Our simulations of the signal and backgrounds in this section
are based upon tree level matrix elements, with the hard-scattering
scale for the structure functions and first-order running
$\alpha_s$ set to the partonic center of mass energy.
For vector-boson plus jet production, we employ 
VECBOS~\cite{Vecbos} running with the structure
functions of Martin, {\it et. al.}~\cite{BCDMS} (the ``BCDMS fit'').
For the processes containing top pairs, we 
perform a Monte Carlo integration of the matrix element
folded with the HMRS(B) structure functions~\cite{MRSEB}.
Under these conditions, the tree-level SM $t \bar t$ production
cross section is 5.1 pb for 163 GeV top quarks,
while two recent computations of the NLO rate including
the effects of multiple soft gluon emission give
$6.95^{+1.07}_{-0.91}$ pb \cite{CERNnlo} 
and $8.12^{+0.12}_{-0.66}$ pb \cite{ANLnlo}
for this mass, implying a $K$ factor in the 1.4 to 1.6 range.
We refrain from applying any
$K$ factor to the rates we report below, although
the reader may wish to do so.  On the the other hand, we do
use a somewhat light value of $m_t$ (163 GeV).

Hadronization and detector effects are mocked up by applying
gaussian smearing with a width of
$125\%/\sqrt{E} \oplus 2.5\%$.  When the simulation of merging jets
is called for, we combine final state partons which lie within
0.4 units of each other in $(\eta,\phi)$ space.
Since our intent is to demonstrate that the superweight method
is viable, we have avoided detailed simulation of the CDF
or D0 detectors.  Instead, we have tried to capture enough
of the general features in order to demonstrate the
viability of the method.  Of course, the superweight criteria 
used in an actual analysis should be determined by the 
experimenters from a complete detector simulation.

\subsection{Discussion of Backgrounds}

There are several ways to mimic our signal of a hard lepton,
missing $E_T$, and two (or more) jets within the Standard Model.
The most obvious background process, and the one with the
largest raw cross section is the direct production of a $W$
plus 2 jets.
However, we can also have
contributions from $Z$ plus 2 jets should one of the leptons
be missed by the detector.
Furthermore, we must beware of Standard Model sources of
top quarks.  In the context of $t \bar t$ production,
the dilepton mode can fake the signal if one of the two leptons
is lost, which is particularly likely if one of the $W$'s decays
to $\tau\nu$.  
Since $\tau$ leptons,
can appear as either a jet of hadrons plus missing momentum
($\tau \rightarrow j \nu_\tau$) or as a lepton plus missing
momentum ($\tau \rightarrow  \ell\bar\nu_\ell\nu_\tau$),
we have been careful to study these backgrounds separately.
The $W$ + 4 jets mode is also a potential troublespot,
since jets can merge or simply be too soft to be detected.
Finally,
single top production followed by SM top decay leads to a
final state of a $W$, two $b$ jets, and missing energy
(plus possibly an extra jet if $W$-gluon fusion is the production
mechanism).  Fortunately, the small rate for single tops is effectively
dealt with by the cuts described below.

The cuts we impose on the data before embarking on our superweight
analysis are listed in Table~\ref{TopToStopCuts}.  
The entries above the dividing line are our ``basic'' cuts.
They were inspired by the CDF top analysis\cite{CDFtop}, in
order to automatically incorporate some of the 
coverage and sensitivity limitations 
imposed by the detector, and to produce a ``clean'' sample
of events.  Thus we require the lepton to have a minimum $p_T$
of 20 GeV, be centrally located ($|\eta|<1$)
and to lie at least 0.4 units in $\Delta R$ from the jets 
($\Delta R \equiv \sqrt{ (\Delta\eta)^2 + (\Delta\varphi)^2}$).
The $p_T$ cut on the lepton aids in the rejection of taus
which decay leptonically.
Some discrimination against events with fake missing $E_T$ is
obtained by setting a minimum 
$\thinspace\thinspace{\not{\negthinspace\negthinspace E}}_T$ of 20 GeV.
The leading two jets should each have a $p_T$ of at least 15 GeV,
and a pseudorapidity $|\eta|<2$.
All jets must have a minimum separation of 0.4 units in $\Delta R$.

To reject Standard Model 
$t \bar{t} \rightarrow W + 4 \enspace{\rm jets}$ events,
we require that the third hardest jet have a {\it maximum}
$p_T$ of 10 GeV.   
While effective in this task, such a cut does have the
unwanted side-effect of suppressing signal events containing
extra jets, such as those containing squarks 
and gluinos~\cite{XtraTops}.
In addition,  some signal events will contain extra jets because of
QCD radiation.   Inclusion of either class of events in the 
data sample requires the relaxation of this cut, as is
done in Sec.~\ref{FullSignal}.   Here we note that the data
in Table~\ref{NevtFULL} imply that no more than 25\% of the
signal events contain extra jets above 10 GeV in $p_T$,
so the ultra-conservative reader may wish to reduce the signals
we report in this section by that amount.  However, since we have
neglected a $K$ factor of 1.4--1.6 in our figures, we feel that
our values are indeed reasonable.

Table~\ref{TopToStopBk} lists the sources of background
discussed above along with the estimated cross section
surviving the cuts for each mode.  
Note that we report the $t\bar{t}$ backgrounds as if
${\cal B}(t\rightarrow Wb)$ were unity:  the actual
contributions to the background in the presence of a
signal are smaller by a factor of this branching ratio squared.
While the basic cuts are nearly adequate for most of the
backgrounds, the contribution
from $W+2$ jets is still an overwhelming 39.1 pb,
necessitating an additional cut.
Given an ideal detector, the 
only source of missing momentum in a background $W$ + 2 jet event
is the neutrino from the decaying $W$.  Hence, 
the transverse mass of the charged lepton and missing momentum
(energy) must be less than or equal to the $W$ mass.
Allowing for the finite width of the $W$ as well as
detector resolution effects, a number of 
events spill over into higher $m_T$ values.  
In contrast, for
SUSY events given by~(\ref{TopToStopSignal}), 
the presence of the two neutralinos in addition to the neutrino
frequently produces events with a transverse mass well above
$m_W$.  Thus, we require that 
$m_T(\ell,\thinspace{\not{\negthinspace p}}_T)>100 {\rm \enspace GeV}$.
This cut is highly effective against the $W$ + 2 jets background, 
while preserving about half of the remaining signal.  
It also removes the small contribution from single top production.
However, it is 
less effective against the $t \bar t$ backgrounds, 
especially those containing $\tau$'s.  Fortunately,
those backgrounds are already under control.

When all of our cuts are imposed, the surviving background
is about 0.42 pb, nearly 90\% of which comes from 
Standard Model production of a $W$ plus 2 jets.
Hence, we consider only that background in developing the
${\cal C}$'s that make up the superweight.

\begin{figure}[h]
\vspace*{15cm}
\includegraphics{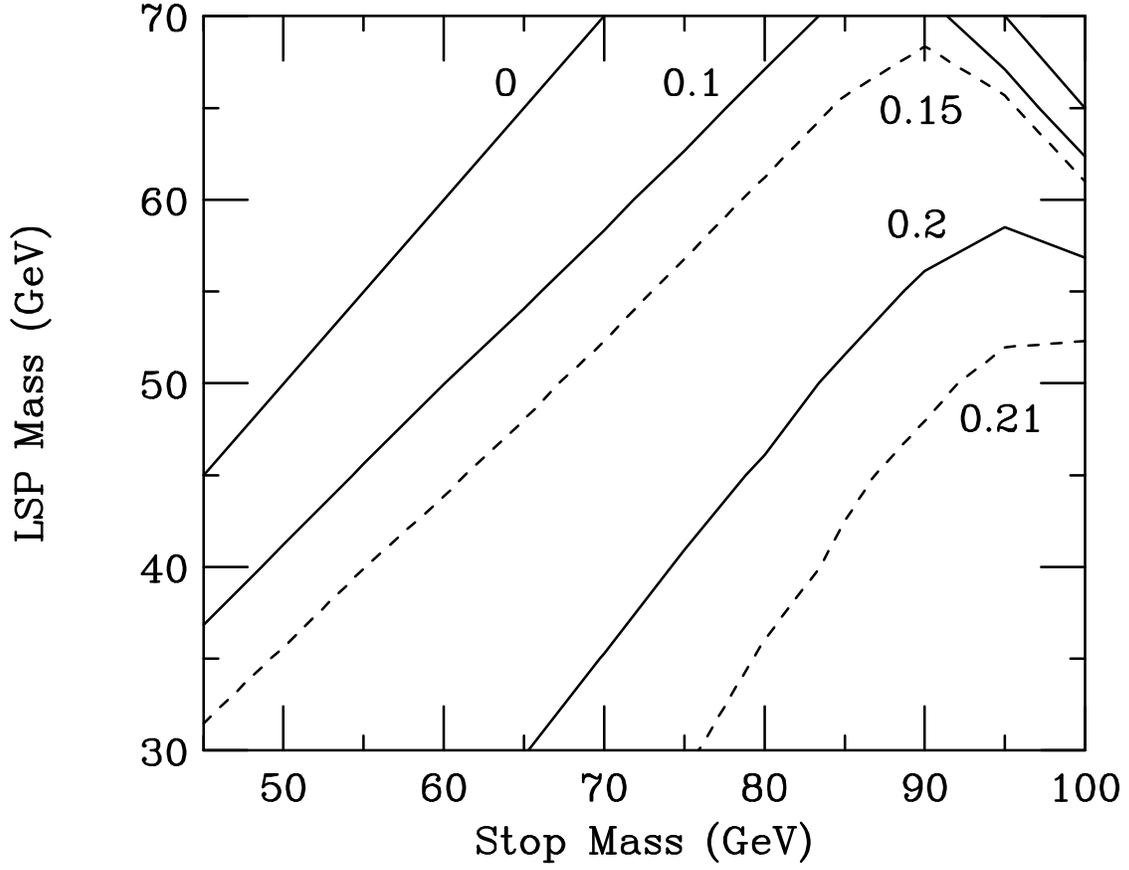}
\vspace{2.0cm}

\caption[]{Fraction of
$t \bar t \rightarrow c \neut1 \neut1 \bar b \ell^{+} \bar\nu_\ell$
events surviving all of the cuts in Table~\ref{TopToStopCuts},
as a function of the stop and LSP
masses.  The efficiency drops to zero at the kinematic limit.
}
\label{EffHi}
\end{figure}

We plot the efficiency for retaining the signal in
Fig.~\ref{EffHi} as a function of the 
stop and LSP masses, and supply numerical values for 
several representative pairings in Table~\ref{TopToStopEff}.

\subsection{Construction of the Superweight}

In Table~\ref{TopToStopSUwgt} we list the 10 criteria used to
build the superweight for the process~(\ref{TopToStopSignal}),
in approximate order of decreasing usefulness.
We now provide intuitive explanations for our selections:
although
the exact placement of the cut points is determined from
the Monte Carlo, we should still be able to understand
from a physical point of view why criteria of the forms
listed are sensible.

We begin our discussion with the three criteria (${\cal C}_5$,
${\cal C}_8$, and ${\cal C}_9$) which depend on joint properties
of the charged lepton ($\ell$) and leading jet ($j_1$).
As already discussed in Sec.~\ref{Features}, the $\bar b$ quark
frequently becomes the leading jet.  Since the $\bar b$ quark
and the charged lepton come from the same parent top quark,
not only would we expect an upper limit on the mass of the
pair (${\cal C}_9$, discussed previously), but there should be some
tendency for the lepton and jet 1 to align.  On the
other hand, in Standard Model $W$ + 2 jet events, the $W$ is recoiling
against the two jets, leading to a tendency for the lepton
and jet 1 to {\it anti-}align.  Hence, we adopt ${\cal C}_5$,
which contributes when the $j_1$-$\ell$ azimuthal angle
is less than 2.4 radians, and ${\cal C}_8$, which contributes
when the cosine of the $j_1$-$\ell$ opening angle is greater
than $-0.15$.

The next group of criteria (${\cal C}_1$, ${\cal C}_2$, 
${\cal C}_6$, ${\cal C}_7$) are various combinations of the
transverse momenta in the event.  Naturally, we make use of
the ``classic'' supersymmetric signature:  the missing 
transverse momentum (${\cal C}_1$), which we require to be
at least 65 GeV to add one unit to the superweight,
that being the point where the two integrated fractions
differ the most.
In addition, we make use
of the fact that Standard Model $W$ + 2 jets production 
falls off rapidly
with increasing $p_T$; that is, we expect the lepton and jets
from the signal process to be somewhat harder on average.
Instead of the
individual $p_T$'s, however, we use their scalar sum with the
missing $p_T$.  Admittedly, there are some correlations introduced
by this choice; however, as discussed in Sec.~\ref{SUwgt},
that is not important for our purposes.

The remaining criteria (${\cal C}_3$, ${\cal C}_4$ and ${\cal C}_{10}$)
may be described as ``miscellaneous.''  
The first of these is tied to the difference between the 
missing $p_T$ and charged lepton $p_T$,
\begin{equation}
\Delta{\cal P}_T \equiv p_T(miss) - p_T(\ell).
\end{equation}
In Standard Model events, 
the neutrino from the decaying $W$-boson
is the only source of missing momentum.  Even though the 2-body
decay of a polarized $W$-boson is not isotropic in its rest frame,
we expect little or no net polarization in the $W$ bosons produced
at the Tevatron.
Consequently, the distribution in $\Delta{\cal P}_T$ 
ought to be symmetric
about zero:  there is no preferred direction for the charged lepton 
relative to the $W$ boost direction.
On the other hand, for events with a supersymmetric origin,
there are a pair of $\neut1$'s in the final state.  On average,
these neutralinos will tend to increase the mean value of 
the missing transverse momentum.  Hence, we expect that the 
distribution in $\Delta{\cal P}_T$ will be 
asymmetric, with a peak for some positive value.
We find that a criterion reading 
$\Delta{\cal P}_T > 0{\rm \enspace GeV}$ is useful.

Earlier, we commented on the use of  
the transverse mass of the charged lepton and missing $p_T$
for the purpose of reducing the $W$ + 2 jets background.
Among the events satisfying this cut, the distributions
{\it still} differ enough to produce a useful superweight criterion:
the spectrum of Standard Model events falls more rapidly than for the
SUSY events.  Thus, we select a criterion of
the form 
$m_T(\ell,\thinspace{\not{\negthinspace p}}_T) > 
125 {\rm \enspace GeV}$ (${\cal C}_4$).

The final criterion we employ is the ``visible'' mass, defined
by summing the observed 4-momenta of the charged lepton and
the leading two jets, and forming an invariant mass-squared.
If all of the final state particles were represented by these
three objects, then this quantity would be equal to the
center of mass energy squared of the hard scattering,
that is $\agt 2{m}_{t}$ for the signal, and $\agt 2M_W$ for
the background.
However,
not all of the particles are detected:  some go down the beampipe,
some are too soft, and some are weakly interacting.
We expect the first two
kinds of losses to be comparable across signal and background.
In contrast, since the signal events contain two extra 
weakly-interacting particles (the $\neut1$'s), an even larger 
proportion of the total mass is invisible.
Although it is not immediately obvious which
way the net effect will go, it is clear that
that distributions in this variable should be different.
From a study like the one described
in Sec.~\ref{SUwgt},  we find that we should set ${\cal C}_{10}=1$ when
$m(\ell,j_1,j_2) < 200 {\rm \enspace GeV}$.

\subsection{Results} 
\label{BasicResults}

The procedure we have in mind for the detection of top to stop
decays is a simple counting experiment.  We apply all of the cuts in 
Table~\ref{TopToStopCuts} to the data, and evaluate the
superweight for each of the surviving events.  Our signal
consists of an excess of events which have a superweight greater
than some value determined by comparing the expected superweight
distributions for the signal and background.

\begin{figure}[h]
\vspace*{15cm}
\includegraphics{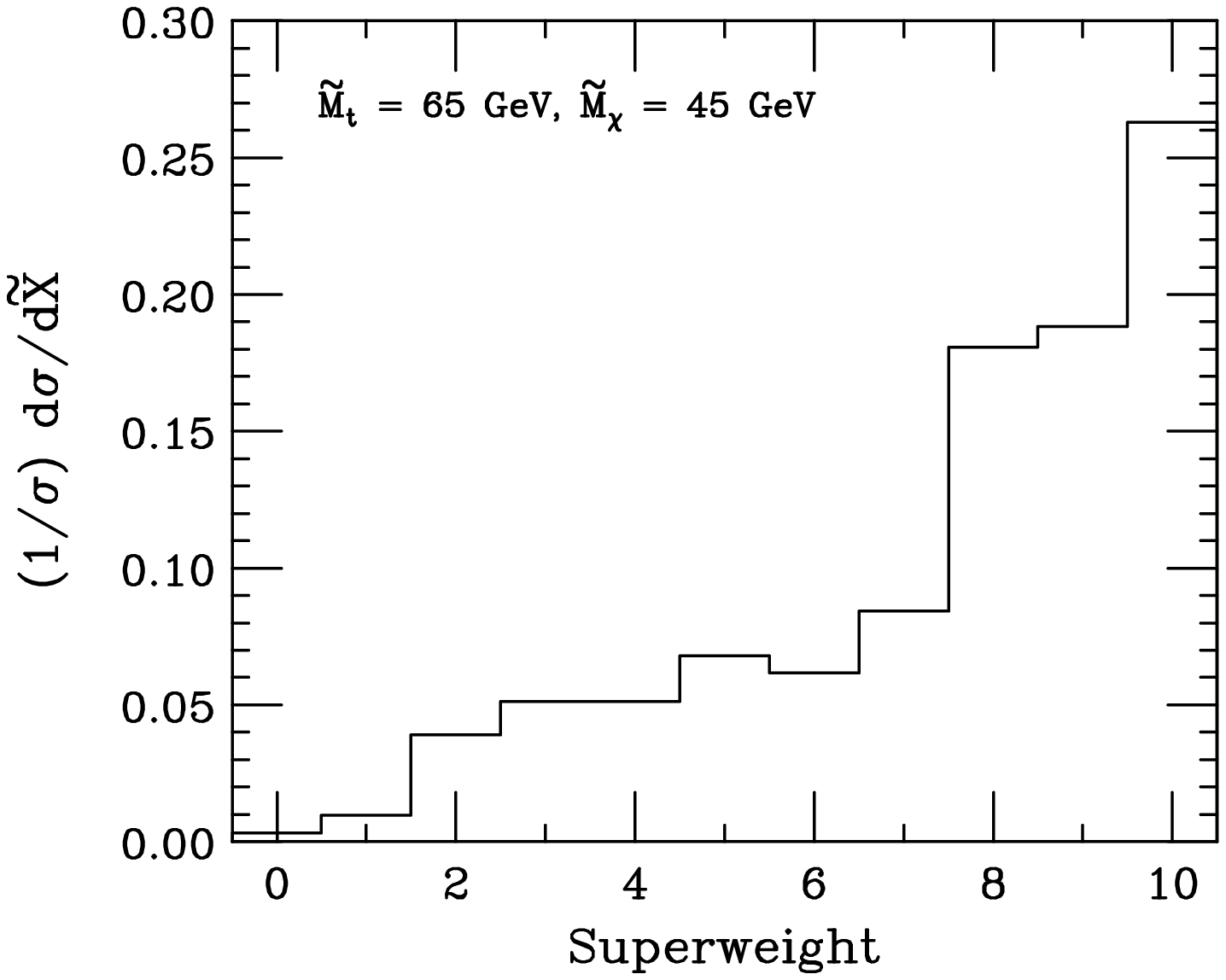}
\vspace{1.0cm}

\caption[]{
Superweight distribution for
$t \bar t \rightarrow c \neut1 \neut1 \bar b \ell^{+} \bar\nu_\ell$
events passing all of the cuts in Table~\ref{TopToStopCuts},
for $\widetilde{M}_{t}=65 {\rm \enspace GeV}$, 
$\mneut1=45 {\rm \enspace GeV}$.
This histogram has been normalized to unit area.
}
\label{SUwgtSigDist}
\end{figure}

\begin{figure}[h]
\vspace*{15cm}
\includegraphics{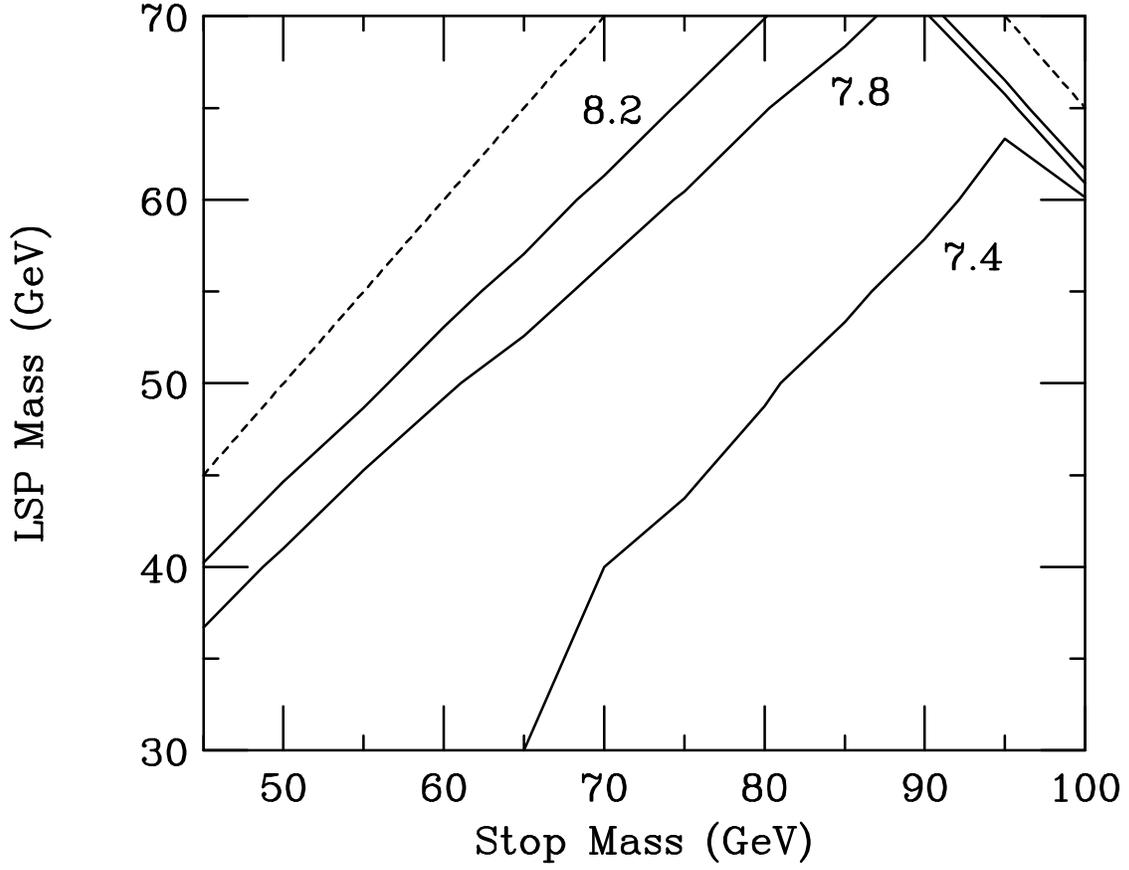}
\vspace{2.0cm}

\caption[]{Mean superweight for 
$t \bar t \rightarrow c \neut1 \neut1 \bar b \ell^{+} \bar\nu_\ell$
events passing all of the cuts in Table~\ref{TopToStopCuts},
as a function of the stop and LSP masses.  
The kinematic limit is indicated by the dotted lines.
}
\label{SUwgtLegoFig}
\end{figure}

We now consider
various pieces of data relevant to evaluating the effectiveness
of the superweight we have just defined.
Fig.~\ref{SUwgtSigDist} shows the distribution of signal events
according to their superweight, for the specific masses
$\widetilde{M}_{t} = 65 {\rm \enspace GeV}$, 
$\mneut1 = 45 {\rm \enspace GeV}$.  A significant
tendency for signal events to have a high superweight is
readily apparent.
Fig.~\ref{SUwgtLegoFig} presents the mean value of $\widetilde{\cal X}$
as a function of the stop and neutralino masses for kinematically
allowed points in the range
$45 {\rm \enspace GeV} \le \widetilde{M}_{t} \le 
100 {\rm \enspace GeV}$, $30 {\rm \enspace GeV} \le 
\mneut1 \le 70 {\rm \enspace GeV}$.
Note the flatness of this distribution:  this implies
that our superweight has roughly the same effectiveness
over the entire range.
Numerical results are presented in 
Table~\ref{TopToStopSUwgtTable} for a few selected points.
Over the entire range the mean superweight is in excess of 7, and
typically 75\% or more of the events have a superweight of 
6 or greater.

\begin{figure}[h]
\vspace*{15cm}
\includegraphics{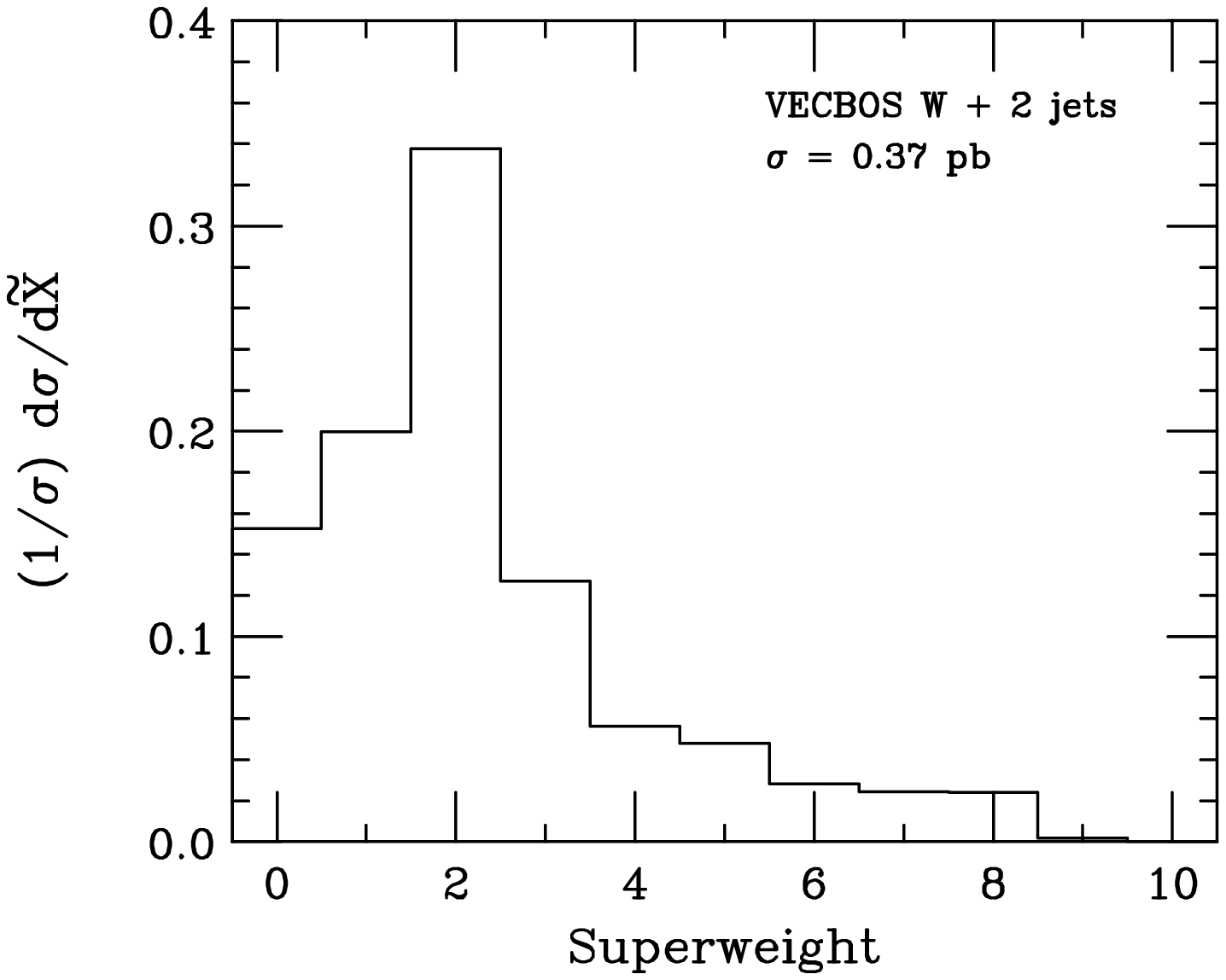}
\includegraphics{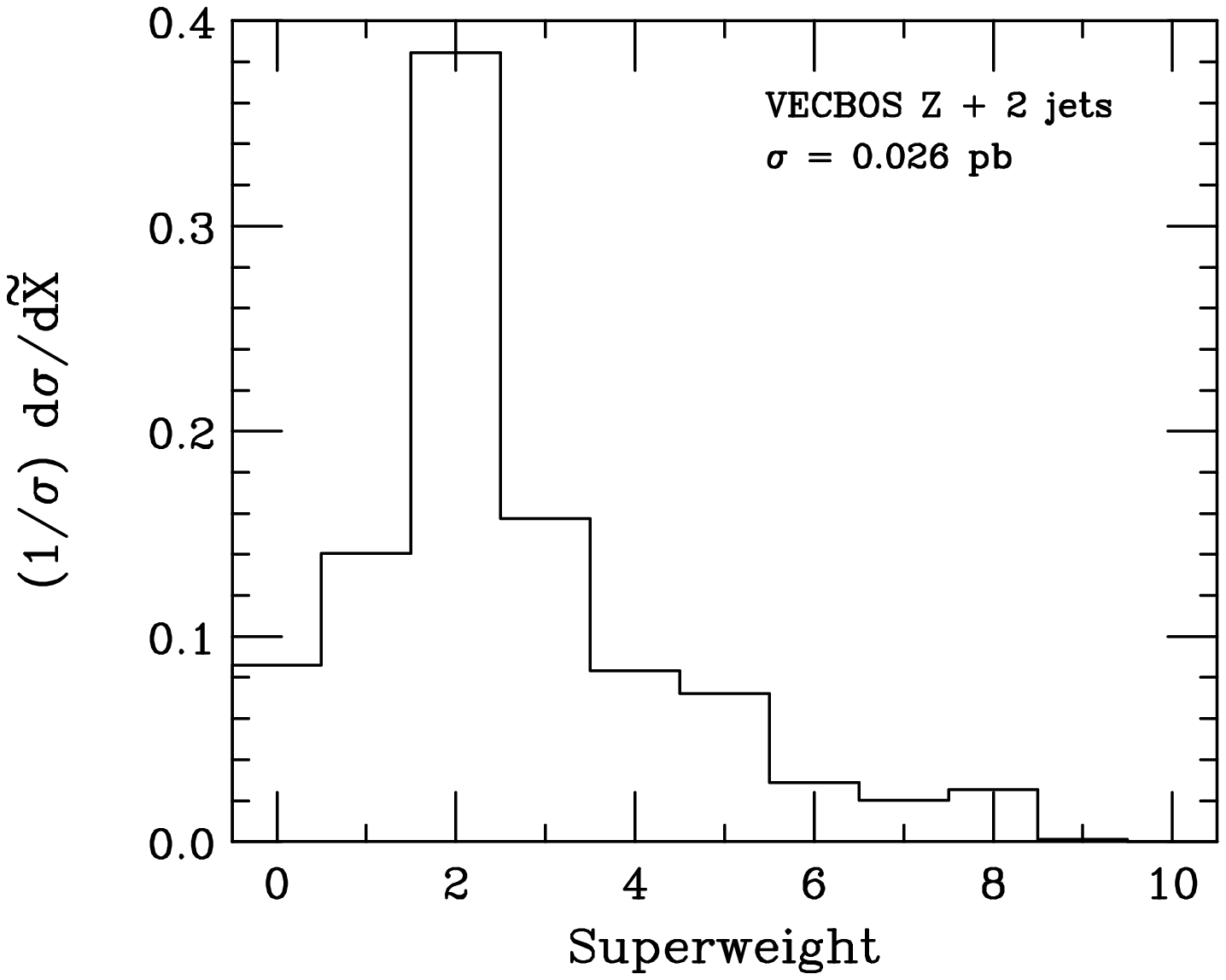}
\includegraphics{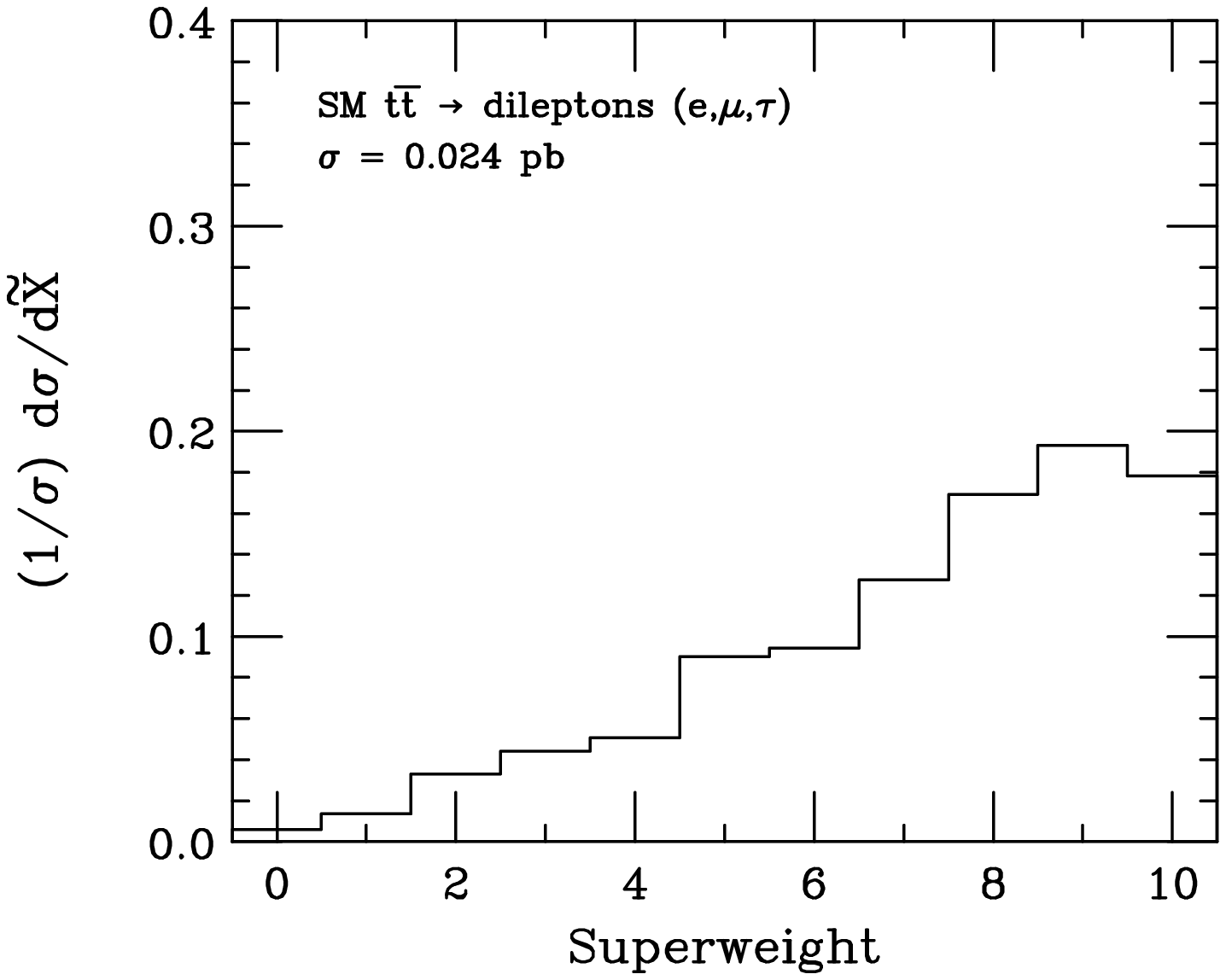}
\includegraphics{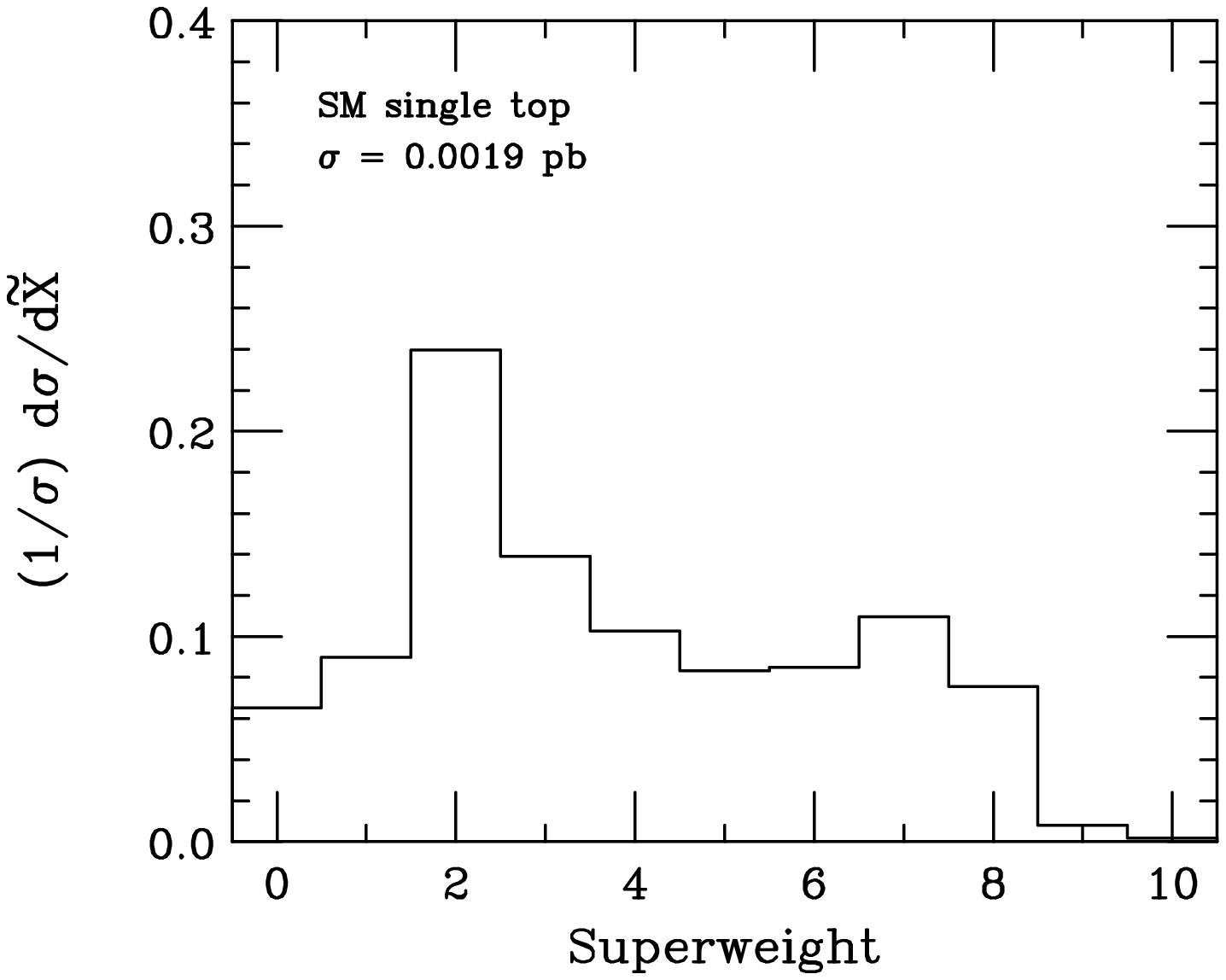}
\vspace{1.0cm}

\caption[]{Superweight distributions for background events
passing all of the cuts in Table~\ref{TopToStopCuts}.
Each histogram has been normalized to unit area.
Upper left: VECBOS $W + 2$ jets (88\% of total).  
Upper right: VECBOS $Z + 2$ jets (6\% of total).  
Lower left: $t\bar{t}\rightarrow{\rm dileptons}$, all 
combinations (6\% of total).
Lower right: single top production ($<$1\% of total).
}
\label{SUwgtBkDist}
\end{figure}

Of course, the significance of these results depends upon the
behavior of the backgrounds.  We plot the superweight distributions
for all backgrounds which were estimated to be 1 fb or greater
in Fig.~\ref{SUwgtBkDist}, and supply the mean values and
fraction of events of each type with superweights of 6 or greater
in Table~\ref{TopToStopBkSUwgtTable}.   It is readily apparent
that our criteria were tailored to reject $W$ + 2 jets events:
they do that very well.   On the other hand, the backgrounds
from Standard Model $t \bar t$ production 
do not typically have low superweights.
In fact, their superweight distributions resemble that of the
signal.  Fortunately, the cross section times branching ratio
surviving our cuts for such events is only 0.023 pb
(0.006 pb if we include the effect of 
${\cal B}(t\rightarrow \tilde{t} \neut1)=50\%$),
while for most (but not all)
values of the SUSY masses, the signal has a cross section 3 to 5
times greater than this particular background.

To get a feeling for the range of masses to which we are
sensitive, we present Fig.~\ref{Nevents}, which shows the
predicted number of signal events in 100 pb$^{-1}$ of data,
the approximate size of the present CDF and D0 data sets.
To guide the eye, we have included the contour where $S/\sqrt{B}=3$.
We must caution the reader, however, that the exact area in which
we can exclude or discover the top squark depends upon a more
complete analysis involving full detector simulations and
Poisson statistics where appropriate.  Note that
the numbers in Fig.~\ref{Nevents} assume a 50\% branching ratio of 
top to stop (which is the most favorable case).  However,
we have omitted the expected increase in rate
from the 1--loop radiative corrections and summation
of multiple soft gluon emission.  Furthermore, we have
reported $t\bar{t}$ backgrounds that do not include the 
effects of the reduced branching ratio to $Wb$.
So overall, we believe our numbers
to be reasonably conservative.  One might hope to increase the
signal somewhat by a careful tuning of the cut choices in 
Table~\ref{TopToStopCuts} and the superweight definition in
Table~\ref{TopToStopSUwgt}.   Also, in the event that a 
signal is found, it would be useful to vary the final cut
on the superweight, as a check on systematics.

It is interesting to compare our results to those of 
Mrenna and Yuan~\cite{Yuan},
who consider the same search, but only employ cuts on the 
``traditional'' event observables.  They obtain a background
of 1.8 events for 100 pb$^{-1}$ of data~\cite{YuanRemark},
compared to our 4.9 events in the $\widetilde{\cal X}\ge6$ sample.
However, the efficiencies they report  for retaining the signal are
only in the 6\%--8\% range:  our efficiencies are as high as 16\%.
The net result is that we have a larger $S/\sqrt{B}$:
indeed, their $S/\sqrt{B}=3$
contour would lie somewhere in the vicinity of the $N=8$ contour on
our Fig.~\ref{Nevents}.

Conspicuously absent from our discussion to this point has
been the issue of $b$-tagging.  We have avoided using 
such information so far for two reasons.
First, the efficiency for $b$-tagging reported by CDF is
currently about 30\% per $b$ jet\cite{TeVMM}.
Hence, the rejection of events without a $b$-tag lowers
the efficiency significantly.  Furthermore, this tagging
efficiency implies that a superweight criterion reading
\begin{equation}
{\cal C} =  \cases{1, & \hbox{if there is a $b$-tag} \cr
                0, & otherwise, \cr}
\end{equation}
only adds about 0.3 units of separation in the mean superweights
of the signal and background.  Compared to the criteria already
in use, this is only a modest separation.  Therefore, we would
prefer to use $b$-tagging to verify that the high superweight
events do indeed contain top quarks in the event that a signal
is observed.  
Note that this assessment would
change should the tagging algorithms improve:
we urge the experimentalists to vary the parameters
and criteria in Tables~\ref{TopToStopCuts} and~\ref{TopToStopSUwgt}
to obtain the optimum balance.
Finally, given that the SUSY signal contains both a 
$b$ jet and a $c$ jet, we remark that the development 
of a specific charm-tagging algorithm would be useful
in this connection.


\section{Inclusion of Squarks and Gluinos} \label{FullSignal}

In this section we consider our superweight analysis in
the context of a ``complete'' SUSY model.  Our aim is to 
demonstrate that the addition of other sources of top quarks
can only help in the observability of a signal, if present.
At the same time, we will show that it is indeed 
sufficient to tune the superweight criteria using the
simplified assumptions of Sec.~\ref{BasicSignal}.
To illustrate these points, we have chosen a specific
model which has a stop mass of 65 GeV and a LSP mass of 45 GeV:
we believe this model to be representative of the
types of models described in Sec.~\ref{Features}.
We list a few other features of this model in Table~\ref{ModelT}.

The data in this section were generated using
{\tt PYTHIA 5.7}~\cite{PYTHIA}
with supersymmetric extensions~\cite{SPYTHIA}.
Tree level matrix elements are used, 
along with the CTEQ2L structure functions~\cite{CTEQ}.
The square of the hard-scattering scale 
for the structure functions and first-order running
$\alpha_s$ is set to the average of the squares of the
transverse masses of the two outgoing particles participating
in the hard scattering (the program default).
For this choice of calculational parameters, the raw SM
$t \bar t$ production cross section is reported as 6.8 pb,
which is rather close to the NLO estimates.
Although we do not do so, the reader may wish to apply 
a $K$ factor of 1.0--1.2 to the signals we report in this
section.

Jets are constructed using a cone algorithm ($R=0.7$) inside a
toy calorimeter using the routine supplied by {\tt PYTHIA}.
No attempt is made to simulate the out-of-cone corrections
required to ensure that the jet energy accurately reflects
the parton energy.  Thus the output systematically underestimates
the jet energies.  As a result, it is not possible to 
directly compare the results appearing in this section
with the results from the previous section.  In particular,
the efficiencies implied by the data in this section will be
lower than what should be expected under actual conditions.
This merely underscores the importance of having each experiment
do the analysis with their full detector simulations in place.
Our goal in this section is to document the effect of
adding  SUSY-induced $t \bar t$ production mechanisms 
to the analysis, and so the only direct comparisons we
need to make to this end are self-contained within this set
of Monte Carlos.

\begin{figure}[h]
\vspace*{15cm}
\includegraphics{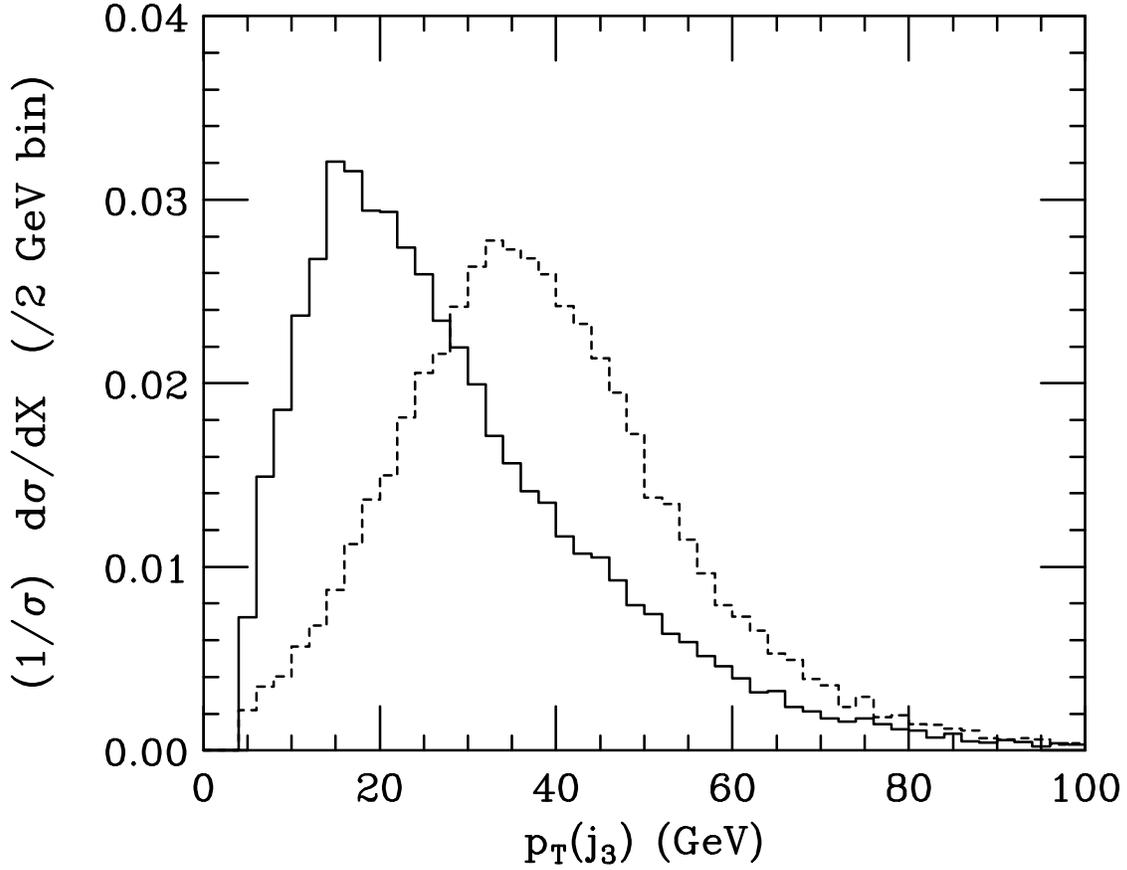}
\vspace{2.0cm}

\caption[]{Transverse momentum distributions of the jet with the
third highest $p_T$ ($j_3$) for 
$\tilde{g}\tilde{g}$, $\tilde{q}\tilde{q}$, and
$\tilde{q}\tilde{g}$ production in the SUSY model summarized in
Table~\ref{ModelT} (solid), and Standard Model
$t\bar{t}\rightarrow W + 4 \enspace\rm{jets}$ (dashed).
Only events which pass all of the cuts in Table~\ref{TopToStopCuts}
(except for the cut on $p_T(j_3)$) are included.
A minimum reconstructed jet energy of 5 GeV is required.
Both histograms have been normalized to unit area.
}
\label{ThirdJet}
\end{figure}

In order to take advantage of the SUSY-produced top events,
we must relax our cut on the $p_T$ of the third jet.
However, we must beware of the background represented
by $W$ + 4 jet SM decays of the top.  In Fig.~\ref{ThirdJet}
we compare the $p_T$ distributions of the third jet for
signal ($\tilde{g}\tilde{g}$, $\tilde{q}\tilde{q}$,
and $\tilde{g}\tilde{q}$) and $t\bar{t} \rightarrow W + 4$ jets 
background, employing the cuts in Table~\ref{TopToStopCuts}
{\it except}\ for the requirement on the third jet.
It is apparent from Fig.~\ref{ThirdJet} that it is possible to 
raise the cut on the maximum allowed $p_T$ of the third jet
without totally swamping the signal in SM $t \bar t$
background.  We will present our results for the cases
$p_T(j_3)<10, 20, 30 {\rm \enspace GeV}$.

Table~\ref{NevtFULL} lists the number of events 
in 100 pb$^{-1}$ predicted to pass the
cuts in Table~\ref{TopToStopCuts}, as a function of the maximum
allowed $p_T$ of jet 3.  The entries above the dividing line
are within the context of our SUSY model.
For the purposes of this study, we define as ``signal'' any
event which contains a pair of $\neut{1}$'s in the final state,
whether or not it contains a $t\rightarrow\tilde{t} c$ decay.
Thus, we list separate entries for $t \bar t$ events which contain
at least one SUSY decay ($t \bar t$ signal) and those which
don't ($t \bar t$ background), but do not distinguish between
squark and gluino events which do or do not contain tops
in the intermediate states.
Should a SUSY scenario of this type prove to be correct
and one wanted to study
only $t\rightarrow\tilde{t} c$ events,
additional work would be required to purify the sample to remove
these non-$t \bar t$ SUSY ``backgrounds.'' 
Note that, as expected,
for the tightest $p_T(j_3)$ cut (10 GeV), the squark and gluino
channels have little effect on the expected number of events.
However, by relaxing this cut to 30 GeV, we allow nearly 2/3
of the $\tilde{g}\tilde{g}$, $\tilde{q}\tilde{q}$, and
$\tilde{g}\tilde{q}$ events into the sample, with only
a modest increase in the background from SM $t\bar{t}$ decays.

\begin{figure}[h]
\vspace*{15cm}
\includegraphics{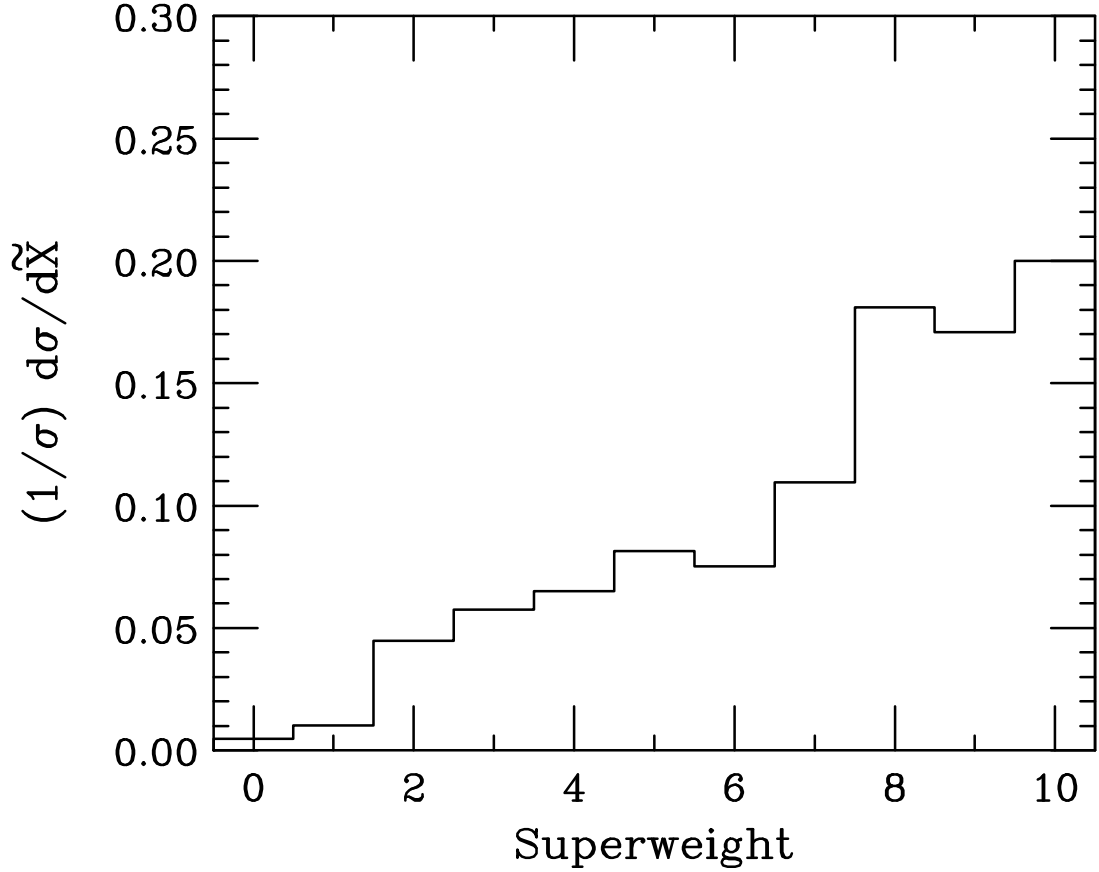}
\vspace{2.0cm}

\caption[]{Superweight distribution for 
$\neut{1}$-containing $t\bar{t}$,
$\tilde{g}\tilde{g}$, $\tilde{q}\tilde{q}$
and $\tilde{g}\tilde{q}$ events in the
SUSY model summarized in Table~\ref{ModelT}
which pass all of the cuts in Table~\ref{TopToStopCuts},
but with a loosened requirement on the third jet, 
$p_T(j_3)<30{\rm \enspace GeV}$.
This histogram has been normalized to unit area.
}
\label{SUwgtFULL}
\end{figure}

The entries below the line give the number of counts assuming
purely SM $t \bar t$ production and decay.  
For good discriminating power, the cuts on $j_3$ and the
superweight should be chosen so that the total number of
counts expected with SUSY is greatly different from the
total number of counts expected without SUSY.  
Since the background from $W/Z + {\rm jets}$ in the absence
of a superweight cut (nearly 40 events) 
is significantly larger than the
entries in Table~\ref{NevtFULL}, it is necessary to impose such a cut.
Fig.~\ref{SUwgtFULL} shows the 
superweight distribution for the signal
($\neut{1}$-containing) events.
Compared to Fig.~\ref{SUwgtSigDist}, we see a somewhat
broader distribution.  However, there is still a significant
peaking at high superweight, and the cut $\widetilde{\cal X}\ge6$ 
still retains the majority of the signal (73\% in this case).
Hence, we present Table~\ref{NevtHI}, which is the same
as Table~\ref{NevtFULL}, but with the additional requirement
$\widetilde{\cal X}\ge6$.  Now the $W/Z + {\rm jets}$ background
is reduced to the point where, for example, taking
the $p_T(j_3)$ cut to be at 30 GeV
yields a factor of 2 difference in the number of 
counts with and without SUSY.
Thus, the prospects for observing or excluding this type of
model are quite good.


\section{Conclusions} \label{CONCLUSIONS}

\begin{figure}[h]
\vspace*{15cm}
\includegraphics{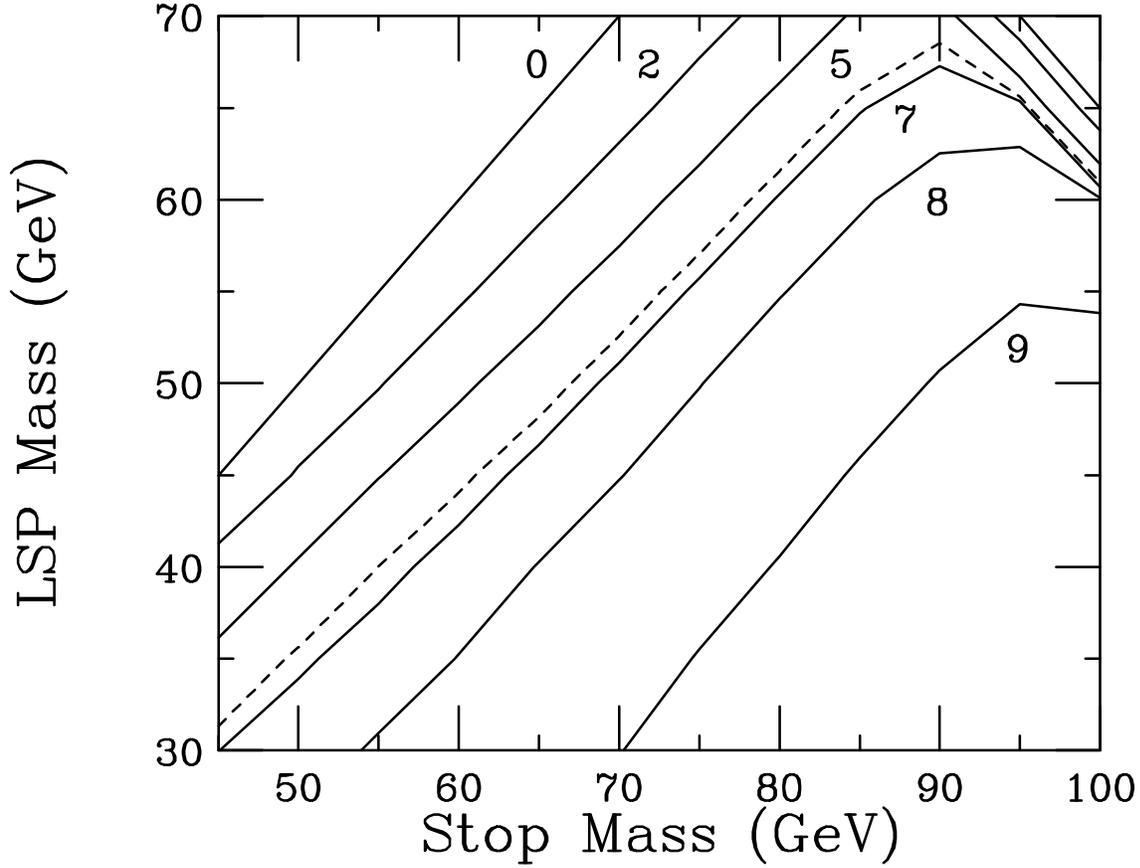}
\vspace{2.0cm}

\caption[]{Predicted number of 
$t \bar t \rightarrow c \neut1 \neut1 \bar b \ell^{+} \bar\nu_\ell$
events in 100 pb$^{-1}$ passing 
all of the cuts in Table~\ref{TopToStopCuts},
and satisfying the condition
$\widetilde{\cal X}\ge6$, as a function
of the stop and LSP masses.  Zero events are predicted 
at the kinematic limit.  
No $K$ factor is included:
see the discussion at the beginning of Sec.~\ref{BasicSignal}.
Table~\ref{TopToStopBkSUwgtTable} indicates that 4.9 background
events are expected for this choice.  The dashed contour marks
the point at which $S/\sqrt{B} = 3$.
}
\label{Nevents}
\end{figure}

We have investigated the possibility of detecting a light
stop squark in the decays of top quarks using the present
Fermilab Tevatron data set (approximately 100 pb$^{-1}$).
Instead of a traditional analysis which relies on cutting
on the kinematic observables individually with a low
resultant efficiency, we have defined a composite 
observable, the superweight.  The superweight is 
assigned event-by-event depending upon how many of the criteria
from a predetermined list are true.  By construction,
events with a large superweight are likely to be signal, 
while those with a small superweight are likely to be
background.  Since we do not require {\it all}\ of the
criterion to be true to accept an event, our efficiency 
is significantly better;  for example, compared to the analysis
of Mrenna and Yuan~\cite{Yuan}, our signal efficiencies are typically
twice as large.  
For the given set of cuts and superweight
criteria, we have shown that the prospects for finding a top-to-stop
signal are good.    Fig.~\ref{Nevents} can be viewed as a summary
of the results.
The collaborations are urged
to view this work as a starting point, since a proper
analysis must be based upon the actual event reconstruction
program used by each experiment.  Furthermore, by adjusting
the parameters in Tables~\ref{TopToStopCuts} and~\ref{TopToStopSUwgt}
it may be possible to do even better.

Finally, we remark that
although we have applied the superweight concept to the specific
case of a light stop squark in supersymmetric models, 
the method is applicable in any situation where the 
individual kinematic cuts
required to reduce the background result in a low signal
efficiency.  Thus, for example, one could consider developing
a superweight suited for the direct search for stop pair production.



\acknowledgements

High energy physics research at the University of Michigan
is supported in part by the U.S. Department of Energy,
under contract DE-FG02-95ER40899.

GDM would like to thank Soo--Bong Kim, Graham Kribs, 
Steve Martin, Steve Mrenna, and Stephen Parke for useful discussions.


\vspace*{1cm}

\begin{table}
\caption{Cuts used for the $t\rightarrow\tilde{t}\neut1$ search.
We use the notation $j$ for 
any of the reconstructed jets, $j_h$ for either of 
the two highest $E_T$
jets, and $j_s$ for any of the additional (soft) jets, if
present.  We refer to the entries above the dividing line
as the ``basic'' cuts. 
\label{TopToStopCuts}}
\begin{tabular}{ccccccccc}
&&&& $p_T(\ell) > 20 {\rm \enspace GeV}$  &&&& \\
&&&& $\thinspace{\not{\negthinspace p}}_T > 
      20 {\rm \enspace GeV}$  &&&& \\
&&&& $p_T(j_h) > 15 {\rm \enspace GeV}$  &&&& \\
&&&& $p_T(j_s) < 10 {\rm \enspace GeV}$  &&&& \\
&&&& $\vert \eta(\ell) \vert < 1 $  &&&& \\
&&&& $\vert \eta(j_h) \vert < 2 $  &&&& \\
&&&& $\Delta R(j,j) > 0.4$  &&&& \\
&&&& $\Delta R(j,\ell) > 0.4$  &&&& \\
\tableline
&&&& $m_T(\ell,\thinspace{\not{\negthinspace p}}_T) 
     > 100 {\rm \enspace GeV} $ &&&&
\end{tabular}
\end{table}


\begin{table}
\caption{Largest backgrounds for the $t\rightarrow\tilde{t}\neut1$ 
search in picobarns surviving the cuts listed in
Table~\ref{TopToStopCuts}.  The values in column I (II) 
include (exclude) the cut on $m_T$.
\label{TopToStopBk}}
\begin{tabular}{ccccddccc}
&&& Background                      &  I     & II      &&& \\[0.1in]
\tableline
&&& $W$ + 2 jets  
                                    & 39.1   &  0.37   &&& \\
&&& $Z$ + 2 jets                    & 0.24   & 0.026   &&& \\
&&& $W$/$Z$ ($\rightarrow$ $\tau\nu / \tau\tau$) + jets   
                            &  2.97  & $<$0.0003 \\
&&& $t \bar t$$\rightarrow$ dileptons  ($ee$, $e
                                       \mu$, $\mu\mu$)\tablenotemark[1]
                                     & 0.014 & 0.0054  &&& \\
&&& $t \bar t$$\rightarrow$ dileptons 
($e\tau$, $\mu\tau$, $\tau\tau$)\tablenotemark[1]
                                     & 0.040 & 0.018 &&& \\
&&& $t \bar t$$\rightarrow W$ + 4 jets\tablenotemark[1]
                                     & 0.0058 & 0.00017 &&& \\
&&& single top\tablenotemark[1]
                                   &  0.10   & 0.0019   &&&
\end{tabular}
\tablenotetext[1] {All top backgrounds assume the absence of
non-SM production, and utilize ${\cal B}(t\rightarrow Wb) = 100\%$.}
\end{table}


\begin{table}
\caption{Cross section times branching ratios in picobarns 
and efficiency  for signal events surviving 
the cuts listed in Table~\ref{TopToStopCuts} for representative
values of $\widetilde{M}_{t}$ and $\mneut1$ in GeV.  
The values in column I (II)
include (exclude) the cut on $m_T$.  The quoted values assume
${\cal B}(t \rightarrow \tilde{t}\neut1)=50\%$.  We do not include
a $K$-factor for radiative corrections: see the 
beginning of Sec.~\ref{BasicSignal}.
\label{TopToStopEff}}
\begin{tabular}{cccccccccc}
&&&$\widetilde{M}_{t}$ & $\mneut1$ &  I  & II &&& \\[0.1in]
\tableline
&&& 60  & 50  &  0.10 (18\%)  & 0.057 (10\%)  &&& \\
&&& 60  & 45  &  0.16 (28\%) & 0.082 (14\%)  &&&  \\
&&& 60  & 35  &  0.21 (38\%) & 0.10 (18\%) &&& \\
&&& 65  & 45  &  0.19 (33\%) & 0.095 (17\%)  &&&  \\
&&& 65  & 35  &  0.23 (40\%) & 0.11 (19\%)  &&& \\
&&& 75  & 45  &  0.23 (40\%) & 0.11 (19\%)  &&& \\
&&& 85  & 50  &  0.24 (42\%) & 0.12 (20\%)  &&& \\
&&& 95  & 60  &  0.23 (41\%) & 0.11 (20\%)  &&& 
\end{tabular}
\end{table}


\begin{table}
\caption{Superweight criteria for $t\rightarrow\tilde{t}\neut1$ search.
One unit is added to the superweight for a given event for each
of the conditions on this list which are satisfied.  Jet 1 refers
to the highest $p_T$ jet in the event, 
and jet 2 to the next-to-highest
$p_T$ jet.  The entries are approximately ordered from most
to least effective.
\label{TopToStopSUwgt}}
\begin{tabular}{ccccccc}
&& Criterion 
    & Quantity         
    & Condition && \\[0.10in]
\tableline
&&${\cal C}_1$ 
    & missing transverse momentum &  
    $\thinspace{\not{\negthinspace p}}_T 
     > 65 {\rm \enspace GeV}$ && \\
&&${\cal C}_2$ 
    & scalar sum of jet 2 and missing $p_T$ & 
    $p_T(j_2) + \thinspace{\not{\negthinspace p}}_T 
     > 95 {\rm \enspace GeV}$ && \\
&&${\cal C}_3$ 
    & difference in missing $p_T$ and lepton $p_T$ &
    $\thinspace{\not{\negthinspace p}}_T - p_T(\ell) 
     > 0 {\rm \enspace GeV}$ \\
&&${\cal C}_4$ 
    & ``$W$'' transverse mass  &  
    $m_T(\ell,\thinspace{\not{\negthinspace p}}_T) 
     > 125 {\rm \enspace GeV}$&& \\
&&${\cal C}_5$ 
    & $j_1$-$\ell$ azimuthal angle & 
    $\varphi_{j_1,\ell} < 2.4$ radians && \\
&&${\cal C}_6$ 
    & scalar sum of charged lepton and missing $p_T$ & 
    $p_T(\ell) + \thinspace{\not{\negthinspace p}}_T 
     > 150 {\rm \enspace GeV}$ && \\
&&${\cal C}_7$ 
    & scalar sum of jet 1 and missing $p_T$ & 
    $p_T(j_1) + \thinspace{\not{\negthinspace p}}_T 
     > 130 {\rm \enspace GeV}$ && \\
&&${\cal C}_8$ 
    & $j_1$-$\ell$ opening angle & 
    $\cos \theta_{j_1,\ell} > -0.15 $ && \\
&&${\cal C}_9$ 
    & $j_1$-$\ell$ transverse mass & 
    $m_T(j_1,\ell) < 125 {\rm \enspace GeV}$ && \\
&&${\cal C}_{10}$
    & visible mass & 
    $m(\ell,j_1,j_2) < 200 {\rm \enspace GeV}$ && 
\end{tabular}
\end{table}


\begin{table}
\caption{Superweight data for
signal events surviving all of
the cuts listed in Table~\ref{TopToStopCuts} for representative
values of $\widetilde{M}_{t}$ and $\mneut1$ in GeV.  
For each pair of values we list
the mean value of the superweight, 
the expected number of events
in 100~pb$^{-1}$ (based upon the entries in Table~\ref{TopToStopEff}),
the expected number of events
with a superweight of 6 (7,8) or greater in 100~pb$^{-1}$.
No $K$ factor is included:
see the discussion at the beginning of Sec.~\ref{BasicSignal}.
\label{TopToStopSUwgtTable}}
\begin{tabular}{cccdddddc}
&$\widetilde{M}_{t}$ & $\mneut1$ &  
     $\langle\widetilde{\cal X}\rangle$ &   
     $N(\widetilde{\cal X}\ge0)$ &
     $N(\widetilde{\cal X}\ge6)$ &
     $N(\widetilde{\cal X}\ge7)$ &
     $N(\widetilde{\cal X}\ge8)$ & \\[0.1in]
\tableline
& 60  & 50  &  7.8 & 5.7 & 4.6 & 4.3 & 3.9 & \\
& 60  & 45  &  7.6 & 8.2 & 6.4 & 5.9 & 5.3 &  \\
& 60  & 35  &  7.4 & 10.4& 8.0 & 7.4 & 6.4 & \\
& 65  & 45  &  7.5 & 9.5 & 7.4 & 6.8 & 6.0 &  \\
& 65  & 35  &  7.4 & 11.0& 8.4 & 7.7 & 6.7 & \\
& 75  & 45  &  7.4 & 11.0& 8.5 & 7.7 & 6.7 & \\
& 85  & 50  &  7.4 & 11.5& 8.8 & 8.0 & 7.0 & \\
& 95  & 60  &  7.4 & 11.2& 8.6 & 7.9 & 6.8 & 
\end{tabular}
\end{table}

\begin{table}
\caption{Superweight data for background events
surviving the cuts listed in
Table~\ref{TopToStopCuts}.   Only those sources with
an expected contribution of 0.001 pb or greater are listed.
For events in each category we list
the mean value of the superweight,
the number of events expected in 100~pb$^{-1}$
(based upon the entries in Table~\ref{TopToStopBk}),
the number of events with a superweight of 6 (7,8) or greater
in 100~pb$^{-1}$.
The bottom line gives the totals, or weighted average, as
appropriate.
\label{TopToStopBkSUwgtTable}}
\begin{tabular}{ccdddddc}
& Background                      &  
     $\langle\widetilde{\cal X}\rangle$ &   
     $N(\widetilde{\cal X}\ge0)$ &
     $N(\widetilde{\cal X}\ge6)$ &
     $N(\widetilde{\cal X}\ge7)$ &
     $N(\widetilde{\cal X}\ge8)$ & \\[0.1in]
\tableline
& $W$ + 2 jets  
                            & 2.3 & 36.5 & 2.9 & 1.8 & 1.0 & \\
& $Z$ + 2 jets             & 2.6 &  2.6 & 0.2 & 0.1 & $<$0.1 & \\
& $t \bar t$$\rightarrow$ dileptons  ($ee$, $e
                                       \mu$, $\mu\mu$)\tablenotemark[1]
                             & 7.1 &  0.5 & 0.4 & 0.3 & 0.3 & \\
& $t \bar t$$\rightarrow$ dileptons  
  ($e\tau$, $\mu\tau$, $\tau\tau$)\tablenotemark[1]
                             & 7.2 &  1.8 & 1.4 & 1.2 & 1.0 & \\
& single top\tablenotemark[1]
                          & 3.8 &  0.2 & $<$0.1 & $<$0.1 & $<$0.1 & \\
\tableline
& Combined                 & 2.6 & 41.6 & 4.9 & 3.6 & 2.3 & \\
\end{tabular}
\tablenotetext[1] {All top backgrounds assume the absence of
non-SM production, and utilize ${\cal B}(t\rightarrow Wb) = 100\%$.}
\end{table}


\begin{table}
\caption{Masses and principle branching ratios for a
SUSY model representative of the scenario discussed 
in Sec.~\ref{Features}.
The input parameters are
$M_1 = 75$, $M_2 = 85$, $M_3 = 200$,
$\mu = -45$ (all in GeV), and $\tan\beta = 1.1$.  The light stop squark
eigenstate
has a mass of 65 GeV and 
${\cal B}( \tilde{t}\rightarrow N_1\thinspace c)=1$.
The heavy stop squark eigenstate
as well as both sbottom squarks have large masses.
The slepton masses in GeV are:  $\widetilde{M}_{e_L} = 115$,
$\widetilde{M}_{e_R} = 125$,
and  $\widetilde{M}_{\nu} = 112$.
The symbol $q$ is used collectively for the $u, d, s, c, b$ quarks.
\label{ModelT}}
\vskip0.35in
\begin{tabular}{cccclrc}
& particle & mass & width & decay modes & B.R. & \\
\tableline
& $t$         & 163 GeV & 2.3 GeV & 
              $t\rightarrow W^{+}b        $ & 52\% & \\
& & & &       $t\rightarrow \neut{1}\thinspace\tilde{t}$ & 33\% & \\
& & & &       $t\rightarrow \neut{3}\thinspace\tilde{t}$ & 12\% & \\
& & & &       $t\rightarrow \neut{2}\thinspace\tilde{t}$ &  3\% & \\
\tableline
&$\tilde{g}$  & 233 GeV & 1.5 GeV &
              $\tilde{g}\rightarrow
                        \tilde{t}^{+}\thinspace\bar{t}$    &  50\% &\\
& & & &       $\tilde{g}\rightarrow
                        \tilde{t}^{-}\thinspace t$         &  50\% &\\
\tableline
&$\tilde{u}_L,\tilde{c}_L$& 259 GeV & 2.9 GeV &
              $\tilde{u}_L \rightarrow
                           \char{i}{+}\thinspace d $ &  49\% &\\
& & & &       $\tilde{u}_L \rightarrow
                           \neut{i}\thinspace u    $ &  26\% &\\
& & & &       $\tilde{u}_L \rightarrow
                           \tilde{g}\thinspace u   $ &  24\% &\\
\tableline
&$\tilde{d}_L,\tilde{s}_L$& 261 GeV & 2.8 GeV &
              $\tilde{d}_L \rightarrow
                           \char{i}{-}\thinspace u $ &  48\% &\\
& & & &       $\tilde{d}_L \rightarrow
                           \tilde{g}\thinspace d   $ &  27\% &\\
& & & &       $\tilde{d}_L \rightarrow
                           \neut{i}\thinspace d    $ &  25\% &\\
\tableline
&$\tilde{u}_R,\tilde{c}_R$& 260 GeV & 1.2 GeV &
              $\tilde{u}_R \rightarrow
                           \tilde{g}\thinspace u   $ &  61\% &\\
& & & &       $\tilde{u}_R \rightarrow
                           \neut{i}\thinspace u    $ &  39\% &\\
\tableline
&$\tilde{d}_R,\tilde{s}_R$& 260 GeV & 1.2 GeV &
              $\tilde{d}_R \rightarrow
                           \tilde{g}\thinspace d   $ &  86\% &\\
& & & &       $\tilde{d}_R \rightarrow
                           \neut{i}\thinspace d    $ &  14\% &\\
\tableline
&$\neut{1}$   &  45 GeV &         &
              STABLE                        &      & \\
\tableline
&$\neut{2}$   &  77 GeV & 0.6 keV & 
              $\neut{2}\rightarrow
                       \neut{1}\thinspace\gamma$       & 82\% &\\
& & & &       $\neut{2}\rightarrow
                       \neut{1}\thinspace q\thinspace\bar{q}$ & 12\% &\\
\tableline
&$\neut{3}$   &  93 GeV & 0.11 MeV& 
              $\neut{3}\rightarrow\neut{1}
                       \thinspace q\thinspace\bar{q}$     & 69\% &\\
& & & &       $\neut{3}\rightarrow\neut{1}
                       \thinspace \nu\thinspace\bar\nu$   & 21\% &\\
& & & &       $\neut{3}\rightarrow\neut{1}
                       \thinspace \ell\thinspace\bar\ell$ & 10\% &\\
\tableline
&$\char{1}{+}$&  83 GeV & 0.41 GeV&
              $\char{1}{+}\rightarrow
                          \tilde{t}^{+}\thinspace\bar{b}$  & 100\% &\\
\tableline
&$\char{2}{+}$& 124 GeV & 0.96 GeV&
              $\char{2}{+}\rightarrow
                          \tilde{t}^{+}\thinspace\bar{b}$  &  93\% &\\
\end{tabular}
\end{table}
\vfill\eject


\begin{table}
\caption{Number of events 
in 100 pb$^{-1}$ predicted to pass 
the cuts in Table~\ref{TopToStopCuts} as a function of
the cut on the third jet, within the context of the SUSY
model described in Table~\ref{ModelT}.
We break SM-produced $t \bar t$ events into
signal and background contributions depending on whether or
not there is a pair of $\neut{1}$'s in the final state.
The third line contains the contributions from all squark and
gluino events, whether or not any top appeared in the intermediate
states.  For comparison, we list the expected number of events
in the absence of SUSY in the last line.   
No $K$ factor is included:  see the discussion at the 
beginning of Sec.~\ref{FullSignal}.
The $W/Z$ + jets background under these
conditions is estimated to be an additional 39.1 events.
\label{NevtFULL}}
\begin{tabular}{ccddddc}
&  & $p_T(j_3)<10{\rm \enspace GeV}$ 
   & $p_T(j_3)<20{\rm \enspace GeV}$ & 
     $p_T(j_3)<30{\rm \enspace GeV}$ & $p_T(j_3)<\infty$ & \\
\tableline
& $t \bar t$ (background)   & 0.9 & 1.4 & 2.2 & 3.7 & \\
& $t \bar t$ (signal)       & 7.9 & 9.4 & 9.8 &10.3 & \\
& $\tilde{g}\tilde{g}$,
 $\tilde{q}\tilde{q}$,
 $\tilde{g}\tilde{q}$    & 0.7 & 2.5 & 4.1 & 6.4 & \\
& total signal           & 8.7 &11.8 &14.0 &16.7 & \\
\tableline
& $t \bar t$ (SM only)      & 3.2 & 5.0 & 7.3 & 13.9 &\\
\end{tabular}
\end{table}

\begin{table}
\caption{Same as Table~\ref{NevtFULL}, but with the additional
requirement $\widetilde{\cal X} \ge 6$.  
The $W/Z$ + jets background under
these conditions is estimated to be an additional 3.1 events.
\label{NevtHI}}
\begin{tabular}{ccddddc}
&  & $p_T(j_3)<10{\rm \enspace GeV}$ 
   & $p_T(j_3)<20{\rm \enspace GeV}$ & 
     $p_T(j_3)<30{\rm \enspace GeV}$ & $p_T(j_3)<\infty$ & \\
\tableline
& $t \bar t$ (background)   & 0.6 & 0.9 & 1.2 & 1.9 & \\
& $t \bar t$ (signal)       & 5.8 & 6.9 & 7.1 & 7.4 & \\
& $\tilde{g}\tilde{g}$,
 $\tilde{q}\tilde{q}$,
 $\tilde{g}\tilde{q}$    & 0.6 & 1.9 & 3.1 & 4.9 & \\
& total signal           & 6.5 & 8.6 &10.3 &12.5 & \\
\tableline
& $t \bar t$ (SM only)      & 2.1 & 3.1 & 4.1 & 6.9 &\\
\end{tabular}
\end{table}


\begin{references}
\bibitem{OtherWay}  We are, of course, free to require instead
${\cal Q} < {\cal Q}_0$ for a non-zero contribution to the superweight.
Our discussion applies to this case as well, with obvious modifications.
\end{references}
\end{document}